# An expert survey to assess the current status and future challenges of energy system analysis

Fabian Scheller*, fjosc@dtu.dk, Energy Economics and System Analysis, Division of Sustainability, Department of Technology, Management and Economics, Technical University of Denmark, Akademivej, 2800 Kgs. Lyngby, Denmark.

Frauke Wiese, frauke.wiese@uni-flensburg.de, Abteilung Energie- und Umweltmanagement, Europa-Universität Flensburg, 24943 Flensburg

Jann Michael Weinand, jann.weinand@kit.edu, Chair of Energy Economics, Institute for Industrial Production (IIP), Karlsruhe Institute of Technology, 76131 Karlsruhe, Germany

Dominik Franjo Dominković, dodo@dtu.dk, Department of Applied Mathematics and Computer Science, Technical University of Denmark, Matematiktorvet, 2800 Kgs. Lyngby, Denmark.

Russell McKenna, russell.mckenna@abdn.ac.uk, Chair of Energy Transition, School of Engineering, University of Aberdeen, King's College, Aberdeen AB24 3FX, United Kingdom & Department of Mechanical and Process Engineering (MAVT), ETH Zurich, 8092 Zürich, Switzerland

* Corresponding author

## Abstract

Decision support systems like computer-aided energy system analysis (ESA) are considered one of the main pillars for developing sustainable and reliable energy transformation strategies. Although today's diverse tools can already support decision-makers in a variety of research questions, further developments are still necessary. Intending to identify opportunities and challenges in the field, we classify modelling capabilities (32), methodologies (15) implementation issues (15) and management issues (7) from an extensive literature review. Based on a quantitative expert survey of energy system modellers (N=61) mainly working with simulation and optimisation models, the status of development and the complexity of realisation of those modelling topics are assessed. While the rated items are considered to be more complex than actually represented, no significant outliers are determinable, showing that there is no consensus about particular aspects of ESA that are lacking development. Nevertheless, a classification of the items in terms of a specially defined modelling strategy matrix identifies capabilities like land-use planning patterns, equity and distributional effects and endogenous technological learning as "low hanging fruits" for enhancement, as well as a large number of complex topics that are already well implemented. The remaining "tough nuts" regarding modelling capabilities include non-energy sector and social behaviour interaction effects. In general, the optimisation and simulation models differ in their respective strengths, justifying the existence of both. While methods were generally rated as quite well developed, combinatorial optimisation approaches, as well as machine learning, are identified as important research methods to be developed further for ESA.

## Highlights:

- Quantitative expert survey about actual modelling characteristics and future realisation complexities.

- Evaluation of modelling capabilities and methodologies, and implementation and management aspects.

- Land-use planning, equity and distributional effect, and technological learning provide modelling potential

- Challenges in modelling non-energy sector and social behavioural interaction effects.

- Data and model documentation standards are considered to be worthy of improvement.

## Contents

# 1. Introduction

A rapid shift to climate neutrality of the global economy is required due to finite fossil resources and the need to limit climate change. In particular, this requires a shift to low-carbon technologies across the energy system, including renewable energy supply and increased efficiency on the demand side. One of the main pillars for supporting global energy transitions involves wide-ranging energy system analysis (ESA) and modelling (ESM) (IRENA 2021). Depending on the structural characteristics of the system under investigation and the purpose of the analysis, the spatial and temporal matching of supply and demand can result in highly complex problems, with a wide range of socio-techno-economic assumptions and different levels of detail at hand (Herbst et al. 2012). This challenge is hardened by the exploitation of renewable energy sources, which require parallel increases in the spatial and temporal resolution of ESMs. In such cases, optimum unit commitments and dispatch for energy systems often cannot be determined analytically but require the use of mathematical optimisation models (Kotzur et al. 2020). The same applies for studying social interactions of imperfectly realistic rational actors on the demand side with the help of agent-based models (Rai und Henry 2016) or for comparing policy measures that differ concerning various key parameters such as costs, emissions, energy supply, and others with simulation models (Lund et al. 2017).

Since assessment findings might be influenced by a wide range of factors, rather holistic frameworks and profound models are needed to sufficiently map the system complexities (Keles et al. 2017). For example, ESMs require sufficient consideration of uncertainties (Yue et al. 2018; Wiese et al. 2018b) or socio-technical factors for improved realisation of optimal transition pathways (Bolwig et al. 2019). Groissböck (2019) demanded a greater detail in terms of modelling ramping of power plants as well as physical and thermodynamic capabilities to not underestimate the complexity of the energy system. Furthermore, Mohammadi et al. (2017) suggest that future multi-generation systems need a broader perspective in terms of energy sources and Hansen et al. (2019) see a greater focus on the joint analysis of multidimensional flexibility options along the energy value chain as important. Ringkjøb et al. (2018) request a better representation of short-term spatiotemporal variability in long-term studies. In cases that high-resolution modelling reaches the limits of being soluble in a reasonable time, the planning and operational step might be divided concerning different time scales with different levels of detail (Pfenninger et al. 2014). Keirstead et al. (2012) call for the utilisation of computational advances like cloud computing for higher complexity modelling, e.g. in terms of activity- and agent-based modelling. However, a higher complexity might also lead to extra data collection efforts due to more specific input parameters (Keirstead et al. 2012) and data uncertainty handling since data quality is important (Keles et al. 2017).

In the past, energy researchers have developed and applied their own models to answer as many research questions as possible (Ridha et al. 2020). Due to a lack of transparency in modelling exercises, for example, about assumptions, data sources, and uncertainties, the energy system modelling field has attracted a lot of criticism (Strachan et al. 2016; DeCarolis et al. 2017). Recently, there have been increasing calls to make ESM and ESA more transparent or even publicly available (Pfenninger et al. 2018; Pfenninger 2017). Nevertheless, existing open-source and commercial energy system models still do not account for all aspects that are necessary for determining successful transition pathways (Groissböck 2019). Moreover, politicians should receive alternative options and recommendations for debating desired energy futures (Lund et al. 2017). A recent study on the trends in tools and approaches of energy system models has also shown that key issues of the models continue to be mainly tool coupling, accessibility and perceived policy relevance (Chang et al. 2021). In the future, therefore, numerous further research challenges will have to be tackled to respond to the future system (Scheller und Bruckner 2019; Wiese et al. 2018b).

Indeed, the strong interest in ESA and ESM in recent decades has inspired many reviews about ESA and ESM to present available tools for different questions, classify model representations and outline future challenges. Building on the classifications, findings and conclusions of 28 review papers of ESA, partially introduced above and systematically compiled in the Supplementary Material A, this paper aims to identify future research opportunities and challenges for ESA. For this, we conducted a quantitative expert survey with a sample size of N=61 in the summer of 2020 to provide insights regarding the criteria *Status of Development* and the *Complexity of Realisation* of 96 identified and classified question items. The compilation and classification of actual representations and future needs from the analysed review papers in terms of modelling capabilities, methodological options, implementation approaches, and management challenges serve as the foundation for our survey. Although there are individual expert-based or rather developer-based surveys for reviewing selected energy modelling tools (Connolly et al. 2010), classifying complexity of ESMs (Ridha et al. 2020), and assessing current trends and challenges in ESA (Chang et al. 2021), our survey-based study employs a broader approach focusing not on representations of specific models but general representations of the research field. Thus, our results enable the identification of key modelling aspects that have been neglected in the past but might be easy to implement in the future, as well as those that will be very challenging. These insights can support researchers, practitioners and policymakers to select suitable focus areas for future research projects.

To achieve this, the paper is structured as follows: the methodology used to create and evaluate the survey is presented in Section 2. Subsequently, our results are presented in Section 3 before the implications and main opportunities and challenges are discussed in Section 4. The paper then concludes in Section 5.

# 2. Methodology

In the present study, future modelling needs are explored with the help of a computer-aided survey. The form of data acquisition is a central challenge in every research project, as it influences survey design, sampling strategy, recruitment procedures, and statistical evaluation techniques. While Section 2.1 presents the data acquisition method and the survey design, Section 2.2 describes the recruitment procedure and section 2.3 the applied evaluation procedure.

## 2.1. Form of data acquisition and design of the survey

As an appropriate technique of data acquisition, a web-based survey was chosen. The structure of the survey was largely determined by the research objective to assess the most urgent improvements and most relevant challenges for ESA. The corresponding question items were systematically derived from a review of various ESM reviews (cf. Section 1) and comprehensive bibliometric analysis of the field (Dominković et al. 2021). An overview of the results of the literature analysis concerning the research scope and future needs is outlined in the Supplementary Material A. During the development process, modelling challenges and future needs were clustered, defined and classified by the research team. The survey started with a selection of socio-demographic questions. Since the participants were asked to answer the questions according to their background information, the model type used and the associated temporal and spatial scale are of particular importance. Subsequently, several challenges needed to be assessed by the respondents. The 96 challenges derived from the synthesis of peer-reviewed energy system analysis reviews were arranged into four sections: 1. Capabilities, 2. Methodology, 3. Implementation, and 4. Management (cf. Table 1). A complete overview of the survey questions is provided in the Supplementary Material B.

In the four survey sections, we employed two main criteria to assess the 96 different items related to different aspects of energy system modelling, namely Status of Development and Complexity of Realisation. While the first criterion Status of Development was related to the question “Which of the following model capabilities would you consider as already represented adequately in the field of energy system analysis and which ones need improvement?”, the second criterion Complexity of Realisation was related to the question “Which of the following model capabilities has been / can be realised without significant difficulties in the field of energy system analysis and which ones not, due to a high level of complexity”. Both criteria were queried on an ordinal scale with a five-point Likert scale ranging from 1-very low to 5-very high.[1] Even though the survey consisted mainly of closed questions, all parts included an option called ‘I don't know’ or ‘not applicable’. Since the questionnaire comprised items derived from the literature body, one open question was asked to the ESM experts to state innovative future modelling directions. In the survey, the term framework was defined as a generic program that can be applied for different use cases (e.g. code and structure), and a model was a corresponding application of a framework (e.g. for a certain set of countries and time resolution including appropriate data). To reduce confusion, we only applied the term “model” in the questions although we are aware that the mentioned challenges could only be tackled on the framework level.

[1] In the management section of the survey (survey section 4) the criterion Complexity of Realization was replaced with Difficulty of Realization. Thereby, the related question was “Which of the following management aspects can be realized without difficulties in the field of energy system analysis and which ones have a high level of difficulty?”.

*Table 1: Composition and structure of the main parts of the computer-aided survey. The challenges to be assessed were arranged into four sections. In this context, sixty-nine items were queried concerning the Status of Development and the Complexity of Realisation. Furthermore, one open question was asked to the ESM experts. The structure of the categories (A, B, C,...) for each section and the numbering of the items (1,2,3,..) is used to describe the results.*

| # | Section title | Section description | Question items |
|---|---|---|---|
| **1** | **Capabilities** | This section deals with the concrete capabilities of models for modelling various relevant aspects of energy systems. To facilitate a detailed analysis, the focus of energy system models typically lies in the simplified techno-economic representation of reality. Relevant model capabilities relate to all parts of the energy value chain. | 8 categories, 32 items;<br>**A) Social aspects and human behaviour modelling:**<br>1. technology acceptance and adoption, 2. lifestyle aspects, 3. stakeholder dynamics and coordination, 4. technology diffusion, 5. equity and distributional effects<br>**B) Demand-side modelling:** 6. Energy service demands, 7. demand-side technology heterogeneity, 8. consumption process models<br>**C) Transmission and distribution system modelling:**<br>9. microgrid and autonomy aspects, 10. power network characteristics, 11. gas network characteristics, 12. heat network characteristics, 13. virtual power plants, 14. ancillary services and spinning reserve<br>**D) Supply generation modelling:**<br>15. ramping capabilities, 16. detailed technology process models, 17. supply-side technology heterogeneity, 18. non-conventional energy supply sources<br>**E) Flexibility, sector coupling and energy system integration modelling:**<br>19. cross-sectoral approaches, 20. multi-energy services and carriers, 21. innovative storage modelling, 22. supply-side flexibility options, 23. demand-side flexibility options<br>**F) Markets and regulations framework modelling:**<br>24. inter-market modelling, 25. market design, 26. regulatory and policy frameworks<br>**G) Environmental and resources modelling:**<br>27. land-use planning patterns, 28. material resource assessments and limitations, 29. nexus issues: for example, land/energy/water/food<br>**H) Feedback and interaction effects:**<br>30. endogenous technology learning, 31. elastic demand, 32. non-energy sector impacts |
| **2** | **Methodology** | This section deals with the methodological approaches of energy system models. Given the analysis purposes and the | 3 categories, 15 items; 1 open question<br>**A) High-resolution modelling:**<br>1. high(er) level of spatial disaggregation, 2. high(er) level of temporal disaggregation, 3. foresight approaches, 4. decomposition methods, 5. soft- or hard-coupling of |
| | | complexity of the system, a variety of approaches have been developed to define and analyse the planning issues, each with its advantages and limitations. Crucial methodological choices are related to the model type, mathematical class, spatial and temporal resolution. | models<br>**B) Programming formulations:**<br>6. new general mathematical frameworks, 7. Non-linear programming formulations, 8. mixed-integer programming formulations, 9. linear programming formulations, 10. stochastic optimisation<br>**C) Model characteristics:**<br>11. consistent and high-quality data sources, 12. higher focus on uncertainty analysis, 13. sustainability indicator assessment, 14. technology neutrality, 15. integrated assessment of multiple capabilities |
| **3** | **Implementation** | This section deals with the implementation of a model, including its usability in general as well as how the model can be applied by different users. Furthermore, documentation standards, as well as technical development options such as standardisation, validation and benchmarking options, affect the implementation options of a model. | 4 categories, 15 items;<br>**A) Development activities:**<br>1. adequate programming language selection, 2. adequate solver selection, 3. modular and adaptable modelling systems, 4. availability of model coupling interfaces<br>**B) Model validation and benchmarking:**<br>5. well-documented model validations, 6. well-documented benchmarks for solving common problems<br>**C) Model usability:**<br>7. (graphical) user interfaces, 8. scenario management tools, 9. web-based and cloud computing environments, 10. master data management systems<br>**D) Documentation standards:**<br>11. installation and application instructions, 12. equation documentations, 13. standards for documentation of data records, 14. clear licensing for code, 15. clear licensing for data |
| **4** | **Management** | This section relates to all operational functions dealing with research projects employing energy system analysis. It covers personnel management, contractual arrangements and can include any functions related to intellectual property, project development, and results dissemination. | 3 categories, 7 items;<br>**A) Human resources management:**<br>1. the possibility of recruiting adequately trained staff, 2. the existence of continuous training<br>**B) Research infrastructure:**<br>3. presence of continuous model maintenance and version control, 4. presence of technical infrastructure<br>**C) Results dissemination:**<br>5. appropriate journals for the publication of the project results, 6. compliance with requirements for open access, open data, and open-source code, 7. public presentation of the project results |

## 2.2. Recruitment procedure

In line with the research objectives, we aimed to obtain responses from leading experts in the field of ESA from all over the world. Potential respondents were identified through the authors' combined networks. Furthermore, key authors in the field were selected in a parallel bibliometric analysis on energy system analysis based on publication numbers and citation indices (Dominković et al. 2021). The web-based survey was created and processed with the aid of the LimeSurvey service provided by Leipzig University. To determine the effectiveness of our survey, a pretest was carried out with experts in our closer networks at the end of June and the beginning of July 2020. The feedback was used to specify the two assessment criteria more precisely and to combine, revise or eliminate question items. While the potential participants were addressed personally, the cover letter of the invitation mail (c.f. Supplementary Material C) explained the intention and background of the study and provided personal contact information. Besides, the study results were offered as an incentive for participation. The addressees received one reminder to participate in the survey. The final sample consisted of 61 completed questionnaires. This corresponds to a response rate of 11%. Fieldwork was completed on 31$^{st}$ October 2020.

An overview of the expert sample working in 23 countries of the world (N=61) in absolute numbers is outlined in Table 2 and Table 3. The respondents' countries cover 70% of the top 20 most productive countries in terms of the total number of publications in the field of ESA (Dominković et al. 2021). Around 75% of the respondents were (senior) researcher or (assistant/ associate/ full) professors. More than half of the respondents were experts working with optimisation models (62%, cf. Figure 1). A further 20% of the respondents were mainly working with simulation models and the final 18% used different types of bottom-up models. In terms of the temporal and spatial scale, the models were distributed quite well between the choices. While around 23% of the models were related to short-term analyses, 30% were related to mid-term analyses, and 47% were related to long-term analyses. Furthermore, 23% focused on a plant or building scale, 41% on a district, municipality or regional scale, and 36% on a national or international scale. In addition, Figure 2 shows the methods used in relation to the working position of the respondents. Most of the respondents worked within public universities (68%) or research institutions (17%) and are male (80%). Around half of the respondents (45%) also reported that they already followed some kind of open source strategy in energy system modelling (fully open data and code: 25%; fully open code but data not or only partly open: 14%; fully open data but code not or only partly open: 7%).

*Table 2: Overview of the survey sample structure. The number of respondents is shown in total and concerning their type of model, they are mainly working with and the institution they are working for. Furthermore, the sex of the respondents is displayed. The survey focuses on bottom-up energy system models: optimisation models (Opt), simulation models (Sim), multi-agent model (Agent), partial equilibrium models (Equil), system dynamics models (Dynm), game-theoretic models (Game), and other bottom-up models (Ors). Due to the low level of respondents and the main focus of this publication, the latter models (*) will be summarised in further analyses. The institutions are abbreviated as follows: university (Uni), research institution (Inst), private company (Comp), public authority (Auth), and others.*

| | **Sample** | **Institution** | | | | | **Gender** | | |
|---|---|---|---|---|---|---|---|---|---|
| | **N** | **Uni** | **Inst** | **Comp** | **Auth** | **Others** | **Male** | **Female** | **Others** |
| Opt | 38 | 26 | 7 | 6 | 1 | 0 | 33 | 4 | 1 |
| Sim | 12 | 12 | 1 | 0 | 0 | 0 | 9 | 3 | 0 |
| Agent* | 1 | 1 | 0 | 0 | 0 | 0 | 1 | 0 | 0 |
| Equil* | 4 | 0 | 2 | 2 | 0 | 0 | 3 | 1 | 0 |
| Dynm* | 2 | 2 | 0 | 0 | 0 | 0 | 2 | 0 | 0 |
| Game* | 1 | 1 | 0 | 0 | 0 | 0 | 0 | 1 | 0 |
| Ors* | 3 | 2 | 1 | 0 | 0 | 1 | 1 | 2 | 0 |
| **Total** | **61** | **44** | **11** | **8** | **1** | **1** | **49** | **11** | **1** |

*Table 3: Overview of the specified model type the respondents are working with. The number of respondents is shown in total and the mode type is furthermore specified with the help of the temporal (from short to long term) and spatial scale (from plant level to international scale). The model types are divided into optimisation models (Opt), simulation models (Sim), multi-agent model (Agent), partial equilibrium models (Equil), system dynamics models (Dynm), game-theoretic models (Game), and other bottom-up models (Ors). Due to the low level of respondents and the main focus of this publication, the latter models (*) will be summarised in further analyses. Thus, there are only three distinct model sub-groups in the actual analysis: optimisation models (Opt), simulation models (Sim) and other models (Agent, Equil, Dynm, Game, Ors).*

| | **Temporal scale** | | | **Spatial scale** | | | | | | |
|---|---|---|---|---|---|---|---|---|---|---|
| | **Short term** | **Mid term** | **Long term** | **Plant level** | **Building scale** | **District scale** | **Munici-pality** | **Region. scale** | **Nation. scale** | **Intern. scale** |
| Opt | 11 | 18 | 31 | 14 | 12 | 12 | 11 | 21 | 27 | 18 |
| Sim | 6 | 5 | 6 | 2 | 4 | 4 | 4 | 5 | 3 | 3 |
| Agent* | 1 | 0 | 1 | 1 | 0 | 0 | 0 | 0 | 1 | 0 |
| Equil* | 0 | 0 | 4 | 0 | 0 | 1 | 2 | 1 | 3 | 3 |
| Dynm* | 1 | 1 | 0 | 1 | 0 | 1 | 1 | 0 | 0 | 0 |
| Game* | 1 | 1 | 0 | 0 | 1 | 1 | 1 | 0 | 0 | 0 |
| Ors* | 1 | 3 | 1 | 2 | 2 | 1 | 2 | 2 | 2 | 2 |
| Total | 21 | 28 | 43 | 20 | 19 | 20 | 21 | 29 | 36 | 26 |

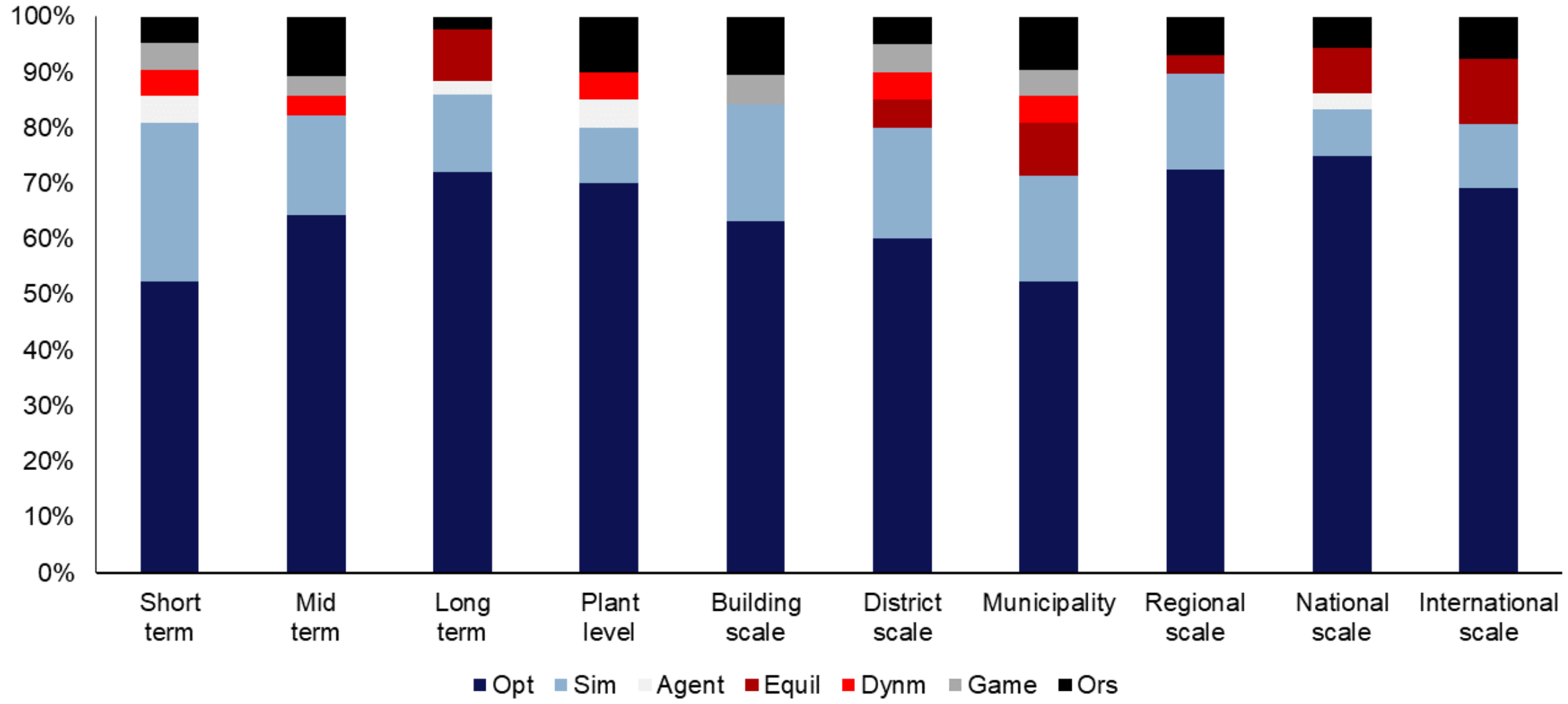

*Figure 1: Overview of the temporal and spatial scales of the models of the sample with simultaneous consideration of the energy system model type (optimisation models (Opt), simulation models (Sim), multi-agent model (Agent), partial equilibrium models (Equil), system dynamics models (Dynm), game-theoretic models (Game), and other bottom-up models (Ors)).*

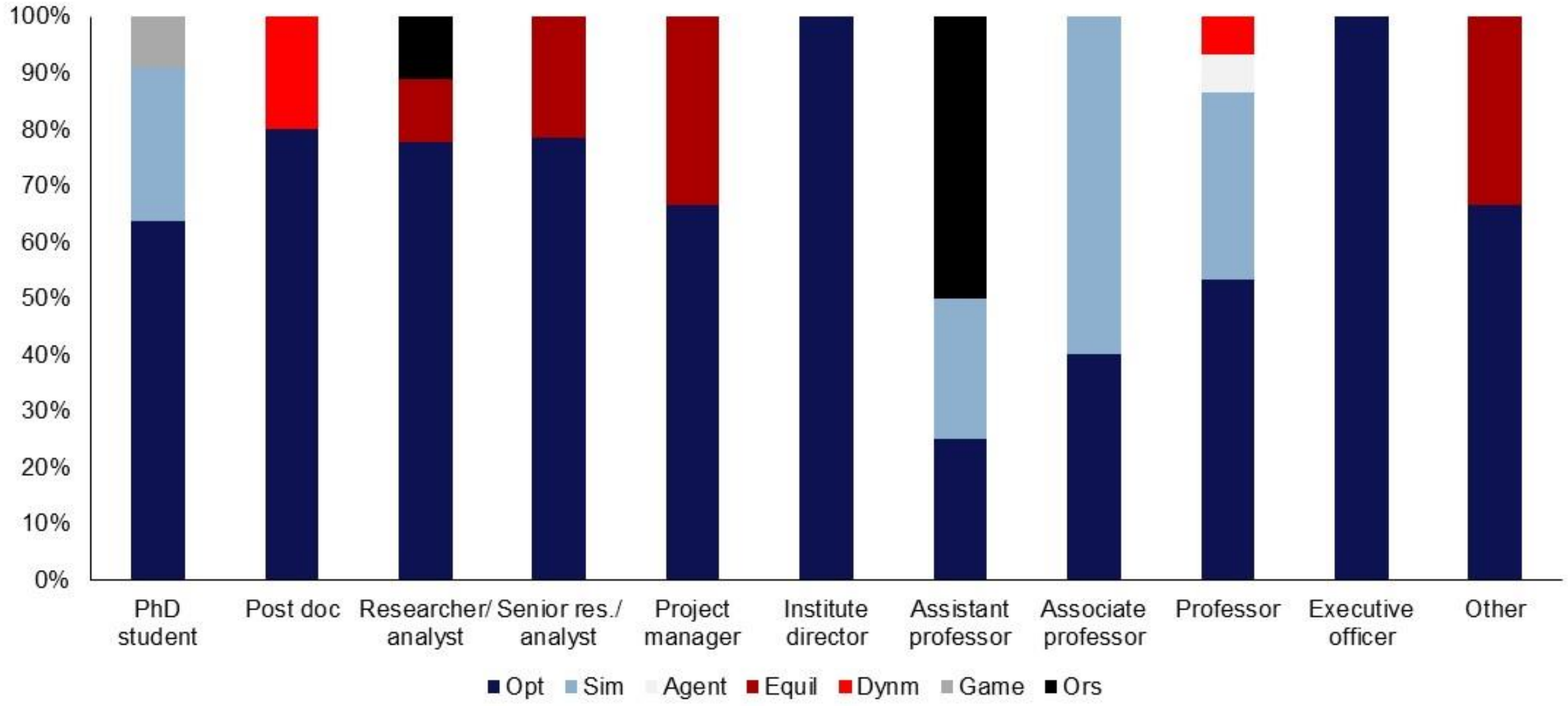

*Figure 2: Overview of the positions of the sample with simultaneous consideration of the energy system model type (optimisation models (Opt), simulation models (Sim), multi-agent model (Agent), partial equilibrium models (Equil), system dynamics models (Dynm), game-theoretic models (Game), and other bottom-up models (Ors)).*

## 2.3. Statistical evaluation procedure

Data analysis was conducted using the software package SPSS (IBM). In this context, four analysis steps were carried out. The order of analysis presented below also corresponds to the structure of the result sub-sections in Section 3.2, Section 3.3, and Section 3.4.

First, the mean ratings of the items regarding the two main criteria Status of Development and Complexity of Realisation were calculated for each survey section. An overview of the results is provided in the Appendix A Tables A1, A2, and A3 from the perspective of the whole survey sample as well as the sub-samples of optimisation model users, simulation model users, and other model users. For a quick orientation, the mean of the rated items across the whole sample is also presented with the help of our specially defined modelling strategy matrix. While an initial overview of all items of all sections is given in Figure 3, a modelling strategy matrix for each section is outlined in Figure 5 and Figure 6, Figure 8, respectively. The 2x2 matrix diagram is a simple square divided into four equal quadrants. Each axis represents a decision criterion, such as Status of Development and Complexity of Realisation. This makes it easy to visualise the items of the different sections that are poorly developed and easy to implement (*low hanging fruits*), poorly developed and complex to implement (*tough nuts*), highly developed and easy to implement (*long runners*), and highly developed and complex to implement (*top stars*). In this context, the mean ratings in tables and matrices allow initial statements about current and future modelling approaches and thus regarding the least and most adequately represented items but also the easiest and most difficult realisations of them.

Second, pairwise Spearman coefficients between the rating of the Status of Development and the complexity of realisation were determined for each of the sub-groups, which provide insights into the interrelations of the main criteria. Thereby, the correlation was defined to be significant at the 10% level. As with the mean ratings, the results are provided in Appendix A Tables A1, A2, and A3.

Third, to underpin the differences between various mean ratings within the distinct sub-groups of respondents who work with optimisation, simulation or other model types, statistically significant relationships between the reported model type and rated criteria were examined using the Kruskal-Wallis H test. The Mann-Whitney U test was then utilised post-hoc to compare each of the identified relationships. Moreover, Mann-Whitney U tests were conducted for other non-distinct sub-groups such as different temporal scales, spatial scales, and respondent's positions. Particular dependencies were highlighted and interpreted in the respective sub-sections. As with the correlation, the 10% confidence level was also used here as the level of significance.

Fourth, a pairwise Spearman correlation matrix of all of the rated items was determined in terms of the whole sample and for the two main assessment criteria. Figure 5 and Figure 7 summarise the coefficients for the modelling capabilities and modelling methodologies. The analysis results allow a description of dependencies between items in one category but also across categories for the two criteria and, thus, answers questions about whether certain items have always been rated similarly. Since there should not be logical dependencies between the items in the last two survey sections (Implementation and Management items respectively), only the first two survey sections (Capability and Methodology items) were taken into account.

Fifth, the answers to the one open question on innovative future modelling directions are discussed and summarised in the discussion.

## 3. Findings

The survey findings are presented individually for each survey section. While Section 3.1 gives an initial insight into the general ratings of all items, the following sections present the rated items in more detail:

Section 3.2 deals with the modelling capabilities, Section 3.3 with the modelling methodology, Section 3.4 with the implementation approach, and Section 3.5 with the project management. The structure of each sub-section follows the analysis steps as described in Section 2.3.

## 3.1 Items overview

The average (mean) ratings across the whole sample of the modelling capabilities, modelling methodologies, and implementation approaches are provided in Figure 3. While we show all the results rounded to one decimal, the vast majority assessed the different items of the different survey sections in a quite similar way without big outliers. The mean overall is 3.0 in terms of the criteria Status of Development and 3.2 in terms of the criteria Complexity of Realisation. Thereby, the items related to the modelling capabilities demonstrate the lowest average (2.9) and the items related to the modelling methodology the highest average (3.2) concerning the assessment criteria Status of Development. In terms of the Complexity of Realisation, the implementation items are rated lowest (3.1) and the methodology items again highest (3.4). At the same time, the standard deviation of the ratings of the Status of Development is higher than the ratings of the Complexity of Realisation for each survey section (Capability: 0.45 vs 0.17; Methodology: 0.37 vs 0.23; Implementation: 0.35 vs 0.34). Thus, there is a lower agreement between the experts on the Status of Development than on the Complexity of Realisation.

Most of the items are assessed as *top stars* according to our modelling strategy matrix, followed by various items which are seen as *tough nuts*. Only a few items are rated as *low hanging fruits* or *long runners*. Since our survey questions are related to well-known modelling aspects, which have been identified with the help of a bibliometric analysis and a literature review as outlined in Section 2.1, these results seem plausible. While there are hardly any outliers, individual capability items are rated lowest in terms of the Status of Development (*lifestyle aspects*, *equity and distributional effects*, *market design*, *non-energy sector impacts*). In contrast, one single methodology item is considered to be the most developed (*linear programming formulations*). Further detailed insights into the assessment of the individual item ratings are given in the following sections.

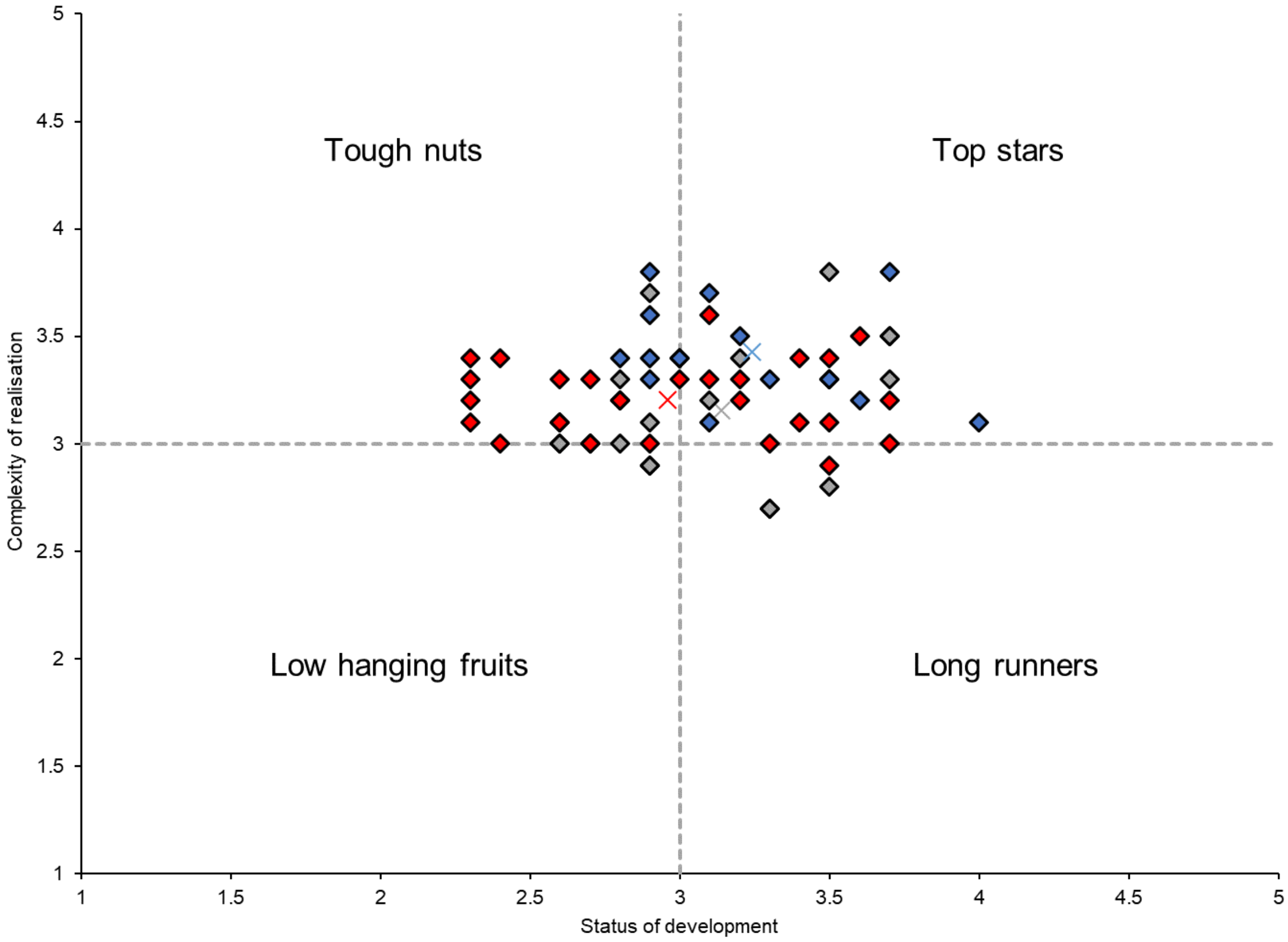

*Figure 3: Overview of the modelling strategy matrix for the average ratings of the modelling capabilities (red), modelling methodologies (blue) and implementation approaches (grey) regarding the Status of Development (ordinal scale very low 1- very high 5) and Complexity of Realisation (ordinal scale very low 1- very high 5) from the perspective of the whole survey sample. The centroids of each item section are depicted with the cross in the respective colour (modelling capabilities (2.9,3.2); modelling methodologies (3.2,3.4); implementation approaches (3.1,3.1)).*

## 3.2 Capability items

The average ratings of the modelling capabilities (cf. Table A1, Figure 4) demonstrate a heterogeneous picture. The two capability items *lifestyle aspects* (2.3±1.0, n=55)[2], as well as e*quity and distributional effects* (2.3±1.0, n=49) of the category social aspects and human behaviour modelling (A), are assessed as the least adequately represented in the field of ESA. With nearly the same average rating, *market design* (2.3±1.2, n=54) and *non-energy sector impacts* (2.3±1.2, n=52) rank close behind them. The associated categories markets and regulations framework modelling (F) and feedback and interaction effects (H) are also the worst represented in today's analyses. In contrast, s*upply-side technology heterogeneity* (3.7±1.1, n=58) and *energy service demands* (3.7±1.2, n=56) are examined to be best represented. The associated categories supply generation modelling (D) and demand-side modelling (B) are also considered to be best represented on the average of all item ratings.

[2] (mean ± standard deviation, sample size)

In terms of the realisation difficulties, all capability items of the category flexibility, sector coupling and energy system integration modelling (E) are assessed with the highest complexity (e.g., d*emand-side flexibility options*: 3.6±1.1, n=60; *cross-sectoral approaches*: 3.5±1.2, n=57). *Technology diffusion* (2.9±1.1, n=51), *micro-grid and autonomy aspects* (3.0±1.3, n=52) and *virtual power plants* (3.0±1.1, n=49) are considered as easiest to implement. It should be noted, however, that the assessments of the Complexity of Realisation are associated with a lower variance. While the differences for the highest and lowest-ranked capabilities is only 0.74 for this criterion, the difference is 1.44 for the Status of Development.

As visualised in the modelling strategy matrix in Figure 4, with moderate realisation complexity and low development status, *land-use planning patterns*, *virtual power plants*, *equity and distributional effects*, and *endogenous technological learning* could represent *low hanging fruits* for future model enhancements. On the other hand, *non-energy sector impacts*, *stakeholder dynamics*, *market designs*, and *lifestyle aspects* are viewed as poorly developed but also complex to implement. These *tough nuts* might represent future features of individual models.

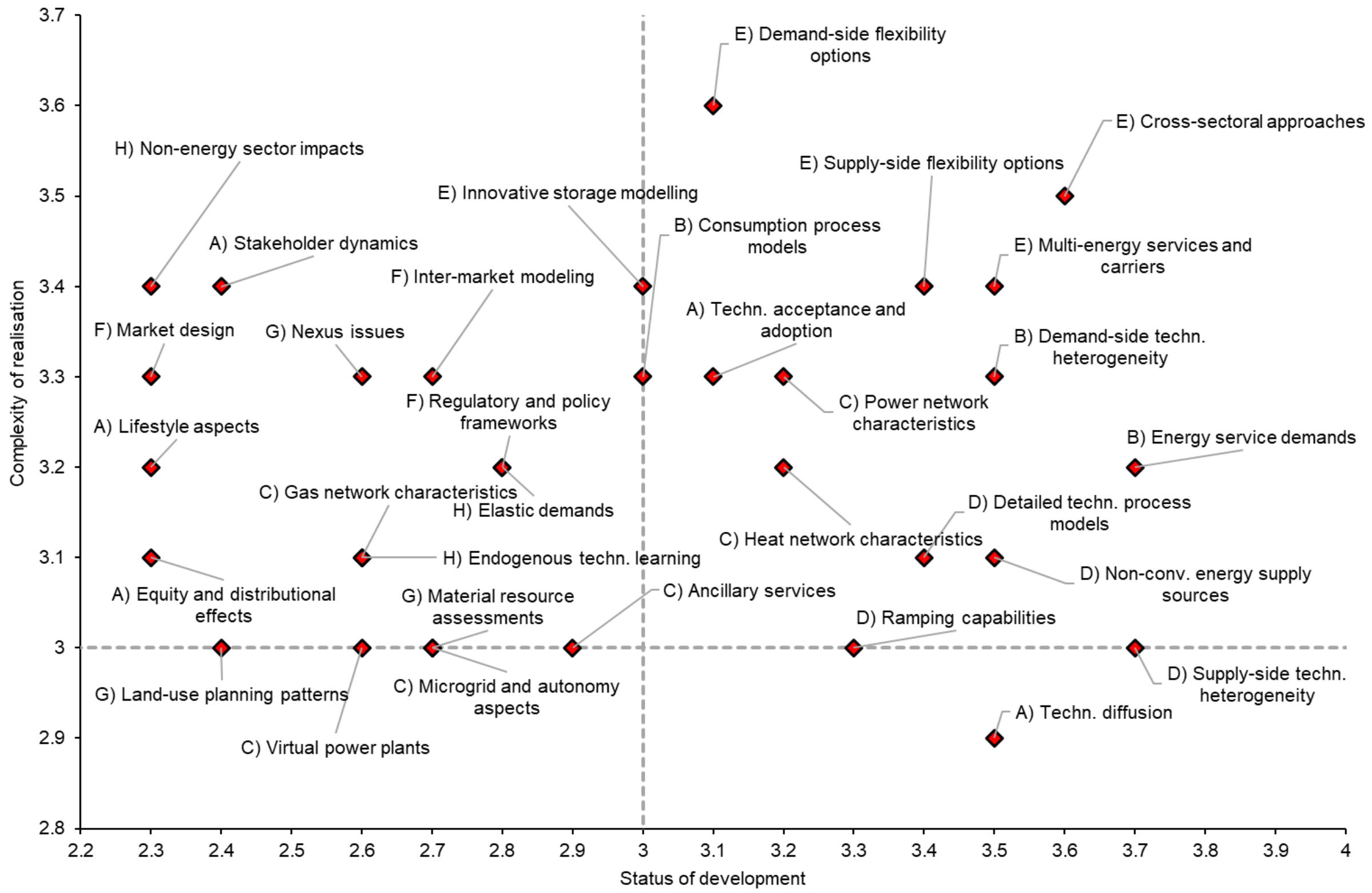

*Figure 4: Overview of the modelling strategy matrix for the average ratings of the modelling capabilities regarding the Status of Development (ordinal scale very low 1- very high 5) and Complexity of Realisation (ordinal scale very low 1- very high 5) from the perspective of the whole survey sample. The categories of the capabilities are social aspects and human behaviour modelling (A), demand-side modelling (B), transmission and distribution system modelling (C), supply generation modelling (D), flexibility,*

*sector coupling and energy system integration modelling (E), markets and regulations framework modelling (F), environmental and resources modelling (G), as well as feedback and interaction effects (H).*

Despite the various mean ratings between the sub-groups of respondents who work with optimisation, simulation or other models (cf. Table A1), statistically significant relationships between the reported model type and rated criteria are found concerning six items for each criterion. While the null hypothesis of the Kruskal–Wallis H test suggests that all the medians are equal, a rejection indicates a statistically significant relationship. In this regard, the null hypothesis is rejected for the items *technology acceptance and adoption*, *ramping capabilities*, *detailed technology process models*, *supply-side flexibility options*, *regulatory and policy frameworks*, and *elastic demand* in terms of the Status of Development. The same is valid in terms of the Complexity of Realisation for the items *technology diffusion*, *multi-energy services and carriers*, *supply-side flexibility options*, *market design*, *endogenous technology learning*, and *elastic demand*. The post-hoc pairwise comparisons of each model type with the Mann–Whitney U test shows that simulation models are more advanced than optimisation models to represent *technology acceptance and adoption*. In contrast, optimisation models are more developed concerning *ramping capabilities* compared with simulation models, detailed *technology process models* compared with simulation models and other models, and *supply-side flexibility options* compared with simulation models. Other models demonstrate a higher development status concerning *regulatory and policy frameworks* and *elastic demands* towards optimisation and simulation models. Besides, unexpectedly the Mann–Whitney U test revealed that all six listed capabilities are indicated as significantly more complex to realise for simulation than optimisation models. Due to the small sample size of individual sub-groups, all comparisons between them, however, need to be treated with caution.

A closer investigation of the impact of different temporal scales (short-term, mid-term, long-term) shows hardly any peculiarities. While the short-term experts report on average a slightly higher Status of Development of capabilities related to the categories transmission and distribution system modelling (C), demand-side modelling (B), and social aspects and human behaviour modelling (A), the long-term experts report a slightly higher Status of Development of capabilities related to the categories markets and regulations framework modelling (F), and flexibility, sector coupling and energy system integration modelling (E). In this context, the capabilities with the highest mean differences are *microgrid and autonomy aspects*, *cross-sectoral approaches*, *ancillary services*, and *consumption process models*. A significant dependency is shown for the items *consumption process models* and *microgrid and autonomy aspects* when we conduct the Mann–Whitney U test from short term and long term perspective. At the same time, the items *cross-sectoral approaches* and *ancillary services* demonstrate only a significant dependency from the perspective of long term modelling or short term modelling, respectively. A similar result is obtained for the items *lifestyle aspects*, *stakeholder dynamics* as well as *regulatory and policy frameworks*, which show a significant relationship with the long term modelling focus towards other responses.

Moreover, a higher average development status rating dependent on the spatial scales (small scale, medium scale, large scale) is indicated for smaller than larger scales for the categories social aspects and human behaviour modelling (A) and transmission and distribution system modelling (C). In contrast, flexibility, sector coupling and energy system integration modelling (E) is rated on average slightly higher by large scale model experts. The capabilities with the highest mean differences are *micro-grid and autonomy aspects*, *lifestyle aspects*, *stakeholder dynamics*, and *cross-sectoral approaches*. This time, the conducted Mann–Whitney U test reveals a significant difference of the ratings for all listed items if one pairwise test the ratings from the small scale but also the large-scale respondents' perspective. Thus, large-scale model experts really rate these items higher. The same is true for e*quity and distributional effects*, *power network characteristics*, and *detailed process models*.

The importance of the capabilities of the same category is again demonstrated with a correlation analysis of the ratings of all capabilities (c.f. Figure 5). Various capabilities of the same category are rated similarly by the total sample. On the one hand, this underlines the focus on certain categories in the past and the importance of other categories in the future. On the other hand, this shows that the capabilities of the same category are conceived together. The interrelations between the modelling capabilities are examined using Spearman's rank correlation coefficient. On the one hand, modelling capabilities of the categories social aspects and human behaviour modelling (A), sector coupling and energy system integration modelling (E), as well as feedback and interaction effects (G) show several significant and strong correlation coefficients regarding the Status of Development and complexity of realisation. On the other hand, capabilities of the environment and resources modelling (F) show a significant and strong correlation coefficient, especially regarding the Complexity of Realisation. In terms of the Status of Development, the strongest significant correlations are between the capabilities *lifestyle aspects* and *equity and distributional effects* (A2, A5: $\rho$=.674), *power network characteristics* and *gas network characteristics* (C10, C11: $\rho$=.663), *ramping capabilities* and *supply-side flexibility options* (D15, E22: $\rho$=.658), *innovative storage modelling* and *supply-side flexibility options* (E21, E22: $\rho$=.657). In terms of the complexity of realisation, the strongest significant correlations are between *market design* and *inter-market modelling* (F24, F25: $\rho$=.837), *nexus issues* and *material resource assessments and limitations* (G28, G29: $\rho$=.761), *regulatory and policy frameworks* and *market design* (F25, F26: $\rho$=.751).

**Complexity of realisation** (columns); **Status of development** (rows)

| | | A | A | A | A | A | B | B | B | C | C | C | C | C | C | D | D | D | D | E | E | E | E | E | F | F | F | G | G | G | H | H | H |
|---|---|---|---|---|---|---|---|---|---|---|---|---|---|---|---|---|---|---|---|---|---|---|---|---|---|---|---|---|---|---|---|---|---|
| | | **1** | **2** | **3** | **4** | **5** | **6** | **7** | **8** | **9** | **10** | **11** | **12** | **13** | **14** | **15** | **16** | **17** | **18** | **19** | **20** | **21** | **22** | **23** | **24** | **25** | **26** | **27** | **28** | **29** | **30** | **31** | **32** |
| **A** | **1** | 1 | .492 | .488 | .450 | .331 | .275 | .338 | .395 | .272 | .285 | .554 | .287 | .455 | .440 | | .297 | | | | | .252 | | .253 | .338 | .317 | | | .284 | .398 | .438 | | .473 |
| **A** | **2** | .345 | 1 | .544 | .315 | .600 | | | .579 | .308 | | .278 | | .366 | .275 | | | | | | | .238 | | .280 | .348 | .362 | .266 | .431 | .392 | .442 | .593 | | .578 |
| **A** | **3** | .547 | .327 | 1 | .302 | .573 | | .279 | .589 | .318 | | .431 | | .442 | .274 | | | | .277 | | | .264 | | | .403 | .579 | .517 | .448 | | .345 | .536 | | .406 |
| **A** | **4** | .526 | .553 | .572 | 1 | .360 | .261 | .304 | .361 | .343 | .247 | | | .363 | .266 | | .335 | | | | .390 | .277 | .252 | | | | | .312 | .321 | .395 | .301 | | |
| **A** | **5** | .378 | .674 | .301 | .500 | 1 | | | .449 | .425 | | .406 | | | .378 | | | | | | | .377 | | | .323 | .486 | | .439 | .308 | .398 | .489 | .404 | .506 |
| **B** | **6** | .487 | | .389 | .352 | | 1 | .588 | .624 | | | .259 | | .312 | | | .277 | .599 | .439 | .426 | .565 | .393 | .353 | .370 | | | | | | | .351 | .441 | |
| **B** | **7** | | | | | | .595 | 1 | .484 | | | | .302 | .500 | .452 | .347 | .279 | .639 | .378 | .476 | .538 | .441 | .442 | .455 | .274 | | .244 | | | | .308 | .462 | |
| **B** | **8** | | .377 | | .374 | .461 | .411 | .541 | 1 | | | .309 | .278 | .480 | .473 | .339 | .320 | .390 | .276 | .307 | .435 | .565 | .426 | .388 | .319 | .305 | .276 | | | | .439 | .436 | .395 |
| **C** | **9** | | | | .504 | .309 | | | .273 | 1 | .517 | .671 | .354 | .454 | .413 | | .246 | | .248 | | | .263 | | .265 | .333 | .456 | .258 | | .415 | .268 | .556 | .251 | .528 |
| **C** | **10** | | | | .294 | | | .292 | .366 | .532 | 1 | .624 | .459 | .440 | .405 | .307 | .333 | | .278 | | | .330 | .346 | .319 | .478 | .471 | .411 | | .298 | | .255 | .242 | .362 |
| **C** | **11** | | .297 | | | .321 | | .346 | .563 | .310 | .663 | 1 | .547 | .541 | .534 | .315 | .299 | | .294 | | .285 | .271 | .325 | .309 | .483 | .660 | .497 | .483 | .382 | .417 | .552 | .324 | .436 |
| **C** | **12** | | | | | | .303 | .400 | .388 | .276 | .541 | .511 | 1 | .348 | .278 | .276 | .307 | .352 | .326 | | .324 | .386 | .287 | .407 | .396 | .408 | .367 | .329 | .371 | .283 | | | .298 |
| **C** | **13** | | | | | | | .321 | | .349 | .420 | .333 | .284 | 1 | .619 | | | .318 | .322 | .294 | .405 | .356 | | .366 | .497 | .566 | .561 | .407 | .376 | .288 | .437 | .289 | .530 |
| **C** | **14** | | | | .402 | | | .399 | .461 | .635 | .462 | .330 | .265 | .469 | 1 | .311 | .258 | | .308 | | .305 | .576 | .335 | .372 | .368 | .399 | | | | | .329 | .473 | .612 |
| **D** | **15** | | | | .264 | | | .377 | .312 | .310 | .449 | .457 | .460 | | .449 | 1 | .625 | .496 | .396 | .322 | .497 | .515 | .610 | .455 | .304 | .325 | .253 | .337 | .368 | .266 | | .516 | |
| **D** | **16** | | | | | | | | | .259 | .302 | | .320 | | .283 | .364 | 1 | .469 | .357 | .257 | .405 | .411 | .485 | .460 | | | | | .384 | .297 | | .334 | |
| **D** | **17** | | | | | | .319 | .418 | .245 | .318 | .333 | .276 | .339 | | .441 | .577 | .598 | 1 | .645 | .567 | .683 | .298 | .514 | .364 | | | | | | | | .405 | |
| **D** | **18** | | | | | | .368 | .379 | | | | | .262 | | .286 | .427 | .545 | .626 | 1 | .513 | .623 | .393 | .545 | .280 | .285 | | .244 | .316 | .416 | .269 | | .506 | .264 |
| **E** | **19** | | | | | | .311 | .420 | | | | | .309 | | | .548 | .250 | .420 | .516 | 1 | .540 | .353 | .701 | .452 | .292 | | .258 | | | | .300 | .499 | |
| **E** | **20** | .271 | | | .263 | | .479 | .492 | .301 | | | | .262 | | .253 | .323 | .363 | .437 | .497 | .631 | 1 | .545 | .655 | .510 | | | .332 | | .285 | .255 | | .647 | |
| **E** | **21** | | | .334 | .474 | | .296 | .413 | .333 | .459 | .357 | .330 | .321 | .333 | .650 | .601 | .299 | .477 | .392 | .596 | .638 | 1 | .577 | .607 | .348 | .373 | .347 | .356 | .424 | | .253 | .577 | .385 |
| **E** | **22** | | | .270 | .385 | | | .436 | .239 | .333 | .372 | .303 | .403 | | .564 | .658 | | .555 | .381 | .512 | .432 | .657 | 1 | .614 | .346 | .252 | .364 | .325 | .395 | .363 | .306 | .680 | .343 |
| **E** | **23** | | | .280 | .463 | | .394 | .558 | .436 | | .336 | .357 | .424 | | .279 | .525 | .359 | .575 | .405 | .513 | .439 | .470 | .611 | 1 | .447 | .384 | .415 | | .479 | .325 | .318 | .558 | .421 |
| **F** | **24** | | | | | | | | | | .272 | .245 | | .295 | .359 | .350 | | .375 | .350 | .390 | | .335 | .370 | .267 | 1 | .837 | .676 | .488 | .609 | .542 | .522 | .442 | .442 |
| **F** | **25** | | | | .320 | | | | | .458 | .412 | .378 | .252 | .294 | | .260 | | | | | | .374 | .264 | | .595 | 1 | .751 | .551 | .563 | .561 | .661 | .490 | .547 |
| **F** | **26** | | | | .337 | | | | | | .265 | .360 | | .274 | | .327 | | .260 | | .337 | | .403 | .332 | .279 | .628 | .622 | 1 | .607 | .471 | .477 | .511 | .450 | .373 |
| **G** | **27** | | .260 | | .276 | | | .333 | | | .246 | .268 | | .352 | .306 | .390 | | .366 | .484 | .497 | .436 | .347 | .315 | .253 | .332 | | .453 | 1 | .608 | .626 | .507 | .353 | .353 |
| **G** | **28** | .433 | | .450 | .462 | .390 | .306 | .260 | | | .270 | .331 | .271 | | | .283 | .490 | .363 | .460 | | .478 | .346 | | .403 | | | .338 | .348 | 1 | .761 | .425 | .461 | .560 |
| **G** | **29** | .409 | | | .317 | .264 | .294 | | | | | | | .317 | | .244 | .460 | .454 | .520 | .290 | .369 | | | .308 | | | | .492 | .609 | 1 | .571 | .412 | .556 |
| **H** | **30** | .407 | .411 | .234 | .392 | .533 | | | .428 | .324 | .257 | .361 | | | .372 | | | | | | | .317 | .322 | .257 | | | | | .311 | .385 | 1 | .461 | .551 |
| **H** | **31** | .267 | .426 | | .419 | .400 | | | .424 | | | .490 | .294 | | .253 | .441 | | .361 | .237 | .286 | .235 | .329 | .317 | .486 | .288 | .256 | .314 | .245 | .436 | .399 | .651 | 1 | .509 |
| **H** | **32** | .428 | | .241 | .357 | .370 | | | | .315 | | .268 | | | | .249 | .424 | .353 | | | .399 | .278 | | .457 | | | | | .506 | .439 | .414 | .592 | 1 |

*Figure 5: Pairwise Spearman correlation matrix (ρ) of the modelling capabilities concerning the Status of Development ratings among each other (lower triangle of the correlation matrix) and the Complexity of Realisation ratings among each other (upper triangle of the correlation matrix). The coefficients are shown for the total sample and only for significant values (correlation is significant at the 10% level). Based on the significant coefficients a colour transition from red over white to blue or rather lower coefficients over medium coefficients to higher coefficients is applied in this table. The categories of the capabilities are social aspects and human behaviour modelling (A), demand-side modelling (B), transmission and distribution system modelling (C), supply generation modelling (D), flexibility, sector coupling and energy system integration modelling (E), environmental and resources modelling (F), feedback and interaction effects (G). The numbers (1-32) are related to the question items of the modelling capabilities (c.f. Table 1).*

## 3.3 Methodology items

Similar to the capabilities, different approaches to the modelling methodology are rated by the experts (c.f. Table A2, Figure 6). On average over the entire sample, the category programming formulations (B) includes the items with the highest as well as the lowest development status. While *new general mathematical framework aspects* (2.8±1.1, n=38) and *Non-Linear Programming* (NLP) formulations (2.9±1.2, n=48) are viewed as underdeveloped, *Mixed Integer Programming* (MIP) (3.7±1.3, n=50) and *Linear Programming* (LP) *formulations* (4.0±1.3, n=55) are considered as highly developed. The last item of the category *Stochastic Optimisation* (SO) formulations (2.9±1.2, n=50) is also seen as not yet fully exploited. In this regard, only the *decomposition methods* (2.9±1.2, n=53) and the focus on *uncertainty analysis* (2.9±1.0, n=59) are rated slightly lower. While there is no clear trend in terms of individual categories, *more complex mathematical structures* are seen as *'tough nuts'* for future models. In terms of the Complexity of Realisation, different capabilities such as *NLP formulations* (3.8±1.1, n=44) or *decomposition methods* (3.6±1.0, n=49), which have been rated rather low in the Status, are considered difficult. At the same time, keeping *consistent and high-quality data sources* is seen throughout as most complex to realise (3.9±1.1, n=59) but also as quite advanced (3.7±1.0, n=60) in the field. Thereby, the ratings regarding the Complexity of Realisation again exhibit a smaller variance as in the previous section.

Modelling methods that might be most suitable for future research in terms of both criteria are *sustainability indicator assessments* and *stochastic optimisation*. At the same time, the modelling strategy matrix does not show any '*low hanging fruits'* (cf. Figure 6). The same is valid for the *long runners*, even though the ratings of the items *LP formulations* and *sustainable indicator assessments* are in a similar range. One reason is the relatively high average rating in terms of Complexity of Realisation. Similar to the capabilities, common items are also assigned a high level of complexity. The various positive Spearman's rank correlation coefficients between the ratings of the Status of Development and the Complexity of Realisation (exception for *LP formulations*; see Table A2) again demonstrate that the higher the respondents rated the development status, the higher they also rated the realisation complexity. This is even more pronounced concerning optimisation models. We might assume that experts who directly work with these approaches experienced a higher problem complexity with each advancement. This is also in line with the fact that more items with low rated status demonstrate a moderate to high and thus significant correlation between the two criteria.

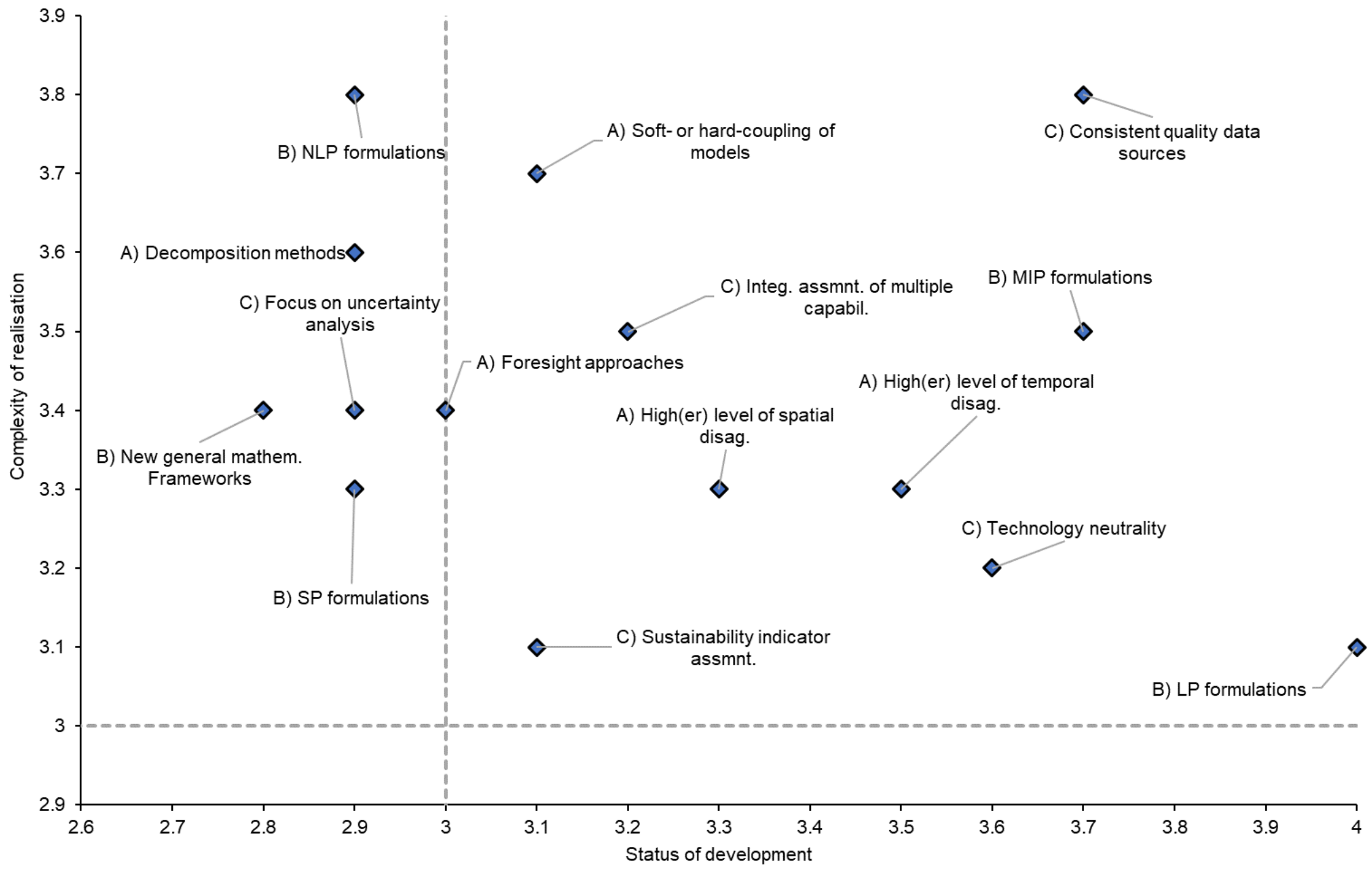

*Figure 6: Overview of the modelling strategy matrix for the average ratings of the modelling methodologies regarding the Status of Development (ordinal scale very low 1- very high 5) and Complexity of Realisation (ordinal scale very low 1- very high 5) from the perspective of the whole survey sample. The categories of the methodologies are high-resolution modelling (A), programming formulations (B), model characteristics (C).*

Regarding the sub-groups and the status of development, the *LP (and MIP) formulations* were rated on average 1.2 (0.8) points higher by the experts of optimisation modelling than the experts of simulation modelling. The same applies to the *assessment of stochastic optimisation* in terms of other models towards the simulation model. Nevertheless, according to the Kruskal–Wallis H test, there are only statistically significant relationships between the reported ratings of *LP formulations* and the sub-groups optimisation and simulation models. Furthermore, the tests also demonstrate a dependency regarding *general mathematical frameworks* for the complexity of realisation. Further pairwise comparisons show that especially the implementation status of *foresight approaches* and *technology neutrality* are assessed significantly different in terms of the spatial and temporal focus of the experts. For realisation complexity, this relationship is demonstrated for both *disaggregation items* as well as *decomposition methods*. From the long term perspective, the experts see a lower complexity level for disaggregation but a higher level for decomposition. From the large scale perspective, the experts report a lower complexity level for the disaggregation and decomposition.

Similar to the capabilities, various methodological items of the same category are rated in the same way (c.f. Figure 7). The analysis with Spearman's rank correlation coefficient show, for example, that experts who report a *high(er) level of spatial disaggregation* also indicate a *high(er) level of temporal disaggregation* for both the status (A1, A2: $\rho$=.582) as well as the complexity (A1, A2: $\rho$=.851). The disaggregation items also demonstrate positive correlations with various other methodology items. This could lead to the assumptions that the spatial and temporal modelling level directly influence the status and complexity perception. Various interrelations are also visible for the status assessment of category B regarding the programming formulations (e.g., B8, B9: ρ=.693 and B8, B10: ρ=.498).

| Status of development | | Complexity of realization | | | | | | | | | | | | | | |
|---|---|---|---|---|---|---|---|---|---|---|---|---|---|---|---|---|
| | | A 1 | A 2 | A 3 | A 4 | A 5 | B 6 | B 7 | B 8 | B 9 | B 10 | C 11 | C 12 | C 13 | C 14 | C 15 |
| A | 1 | 1 | .851 | .624 | .396 | .322 | .542 | .492 | .332 | | .608 | .425 | .462 | .375 | .287 | .375 |
| A | 2 | .582 | 1 | .501 | .420 | .323 | .568 | .474 | .350 | .241 | .455 | .368 | .467 | .329 | .401 | .342 |
| A | 3 | .346 | .385 | 1 | .303 | .343 | .417 | .379 | | | .599 | .253 | .644 | .410 | | .407 |
| A | 4 | .410 | .432 | .485 | 1 | .424 | .553 | .442 | | | .534 | | .321 | | .320 | .593 |
| A | 5 | .346 | .397 | .401 | .535 | 1 | | .286 | .263 | | | .278 | .254 | | | .267 |
| B | 6 | .405 | .470 | | .362 | | 1 | .554 | .298 | | .649 | | .546 | | | |
| B | 7 | | | .320 | .465 | .258 | .490 | 1 | .372 | | .586 | | .399 | .371 | | .430 |
| B | 8 | .364 | .447 | .407 | .312 | .381 | .400 | .352 | 1 | .337 | .401 | .341 | | | | |
| B | 9 | .397 | .347 | .252 | .326 | .433 | .386 | | .693 | 1 | | | | | .480 | .333 |
| B | 10 | .520 | .440 | .390 | .459 | .394 | .366 | .315 | .498 | .367 | 1 | .281 | .572 | .275 | | .330 |
| C | 11 | | .401 | .247 | | .387 | | | | | | 1 | .398 | | .357 | .266 |
| C | 12 | .403 | .257 | .345 | .307 | .435 | | .365 | .326 | .462 | .374 | .334 | 1 | .569 | | .378 |
| C | 13 | | | | | .335 | | | .306 | | .262 | | 0.44 | 1 | .337 | .515 |
| C | 14 | .287 | .371 | .487 | | .466 | | | | .366 | | .489 | .370 | | 1 | .522 |
| C | 15 | | .248 | .368 | .498 | .485 | | .482 | .306 | .381 | | .310 | .482 | .369 | .433 | 1 |

*Figure 7: Pairwise Spearman correlation matrix (ρ) of the modelling methodologies concerning the Status of Development ratings among each other (lower triangle of the correlation matrix) and the Complexity of Realisation ratings among each other (upper triangle of the correlation matrix). The coefficients are shown for the total sample and only for significant values (correlation is significant at the 10% level). Based on the significant coefficients a colour transition from red over white to blue or rather lower coefficients over medium coefficients to higher coefficients is applied in this table. The categories of the methodologies are high-resolution modelling (A), programming formulations (B), model characteristics (C). The numbers (1-15) are related to the question items of the methodological approaches (c.f. Table 1).*

## 3.4 Implementation items

The ranked implementation approaches (c.f. Table A3, Figure 8) show the lowest status for three items of the category model usability (C) on average. While *web-based and cloud environments* are considered as worst represented in today's modelling systems (2.6±1.2, n=51), *master data management systems* (2.8±1.4, n=51) and *graphical user interfaces* (2.8±1.3, n=54) are similarly viewed as underrepresented. This is followed by documentation standards (D). Thereby, *clear licensing for data* (2.9±1.2, n=49) and *standards for documentation of data records* (2.9±1.2, n=56) are seen as important. The other items of the category *equations documentation* (3.5±1.0, n=57) and *clear licensing for code* are, in contrast, considered to be better developed (3.3±1.2, n=50). Only two items are even rated higher than *equations documentation* within the criterion Status of Development: *adequate solver* (3.7±1.0, n=55) and *adequate programming language selection* (3.7±1.0, n=57). Both belong to the category development activity (A). Despite the high-status ratings of the items in category (A), the experts throughout recognise their complexity. This might be traced back to strong activity in the field over the past decade(s). *Modular and adaptable modelling systems* (3.8±1.0, n=57), *availability of model coupling interfaces* (3.7±1.0, n=55), and *adequate solver selection* (3.5±1.0, n=52) are rated on average as the most complex aspects to realise. On the other hand, e.g. web-based and cloud computing environments, graphical user interfaces, clear licensing for data and standards for data documentation are seen as '*low hanging fruits'* according to our Modelling Strategy Matrix.

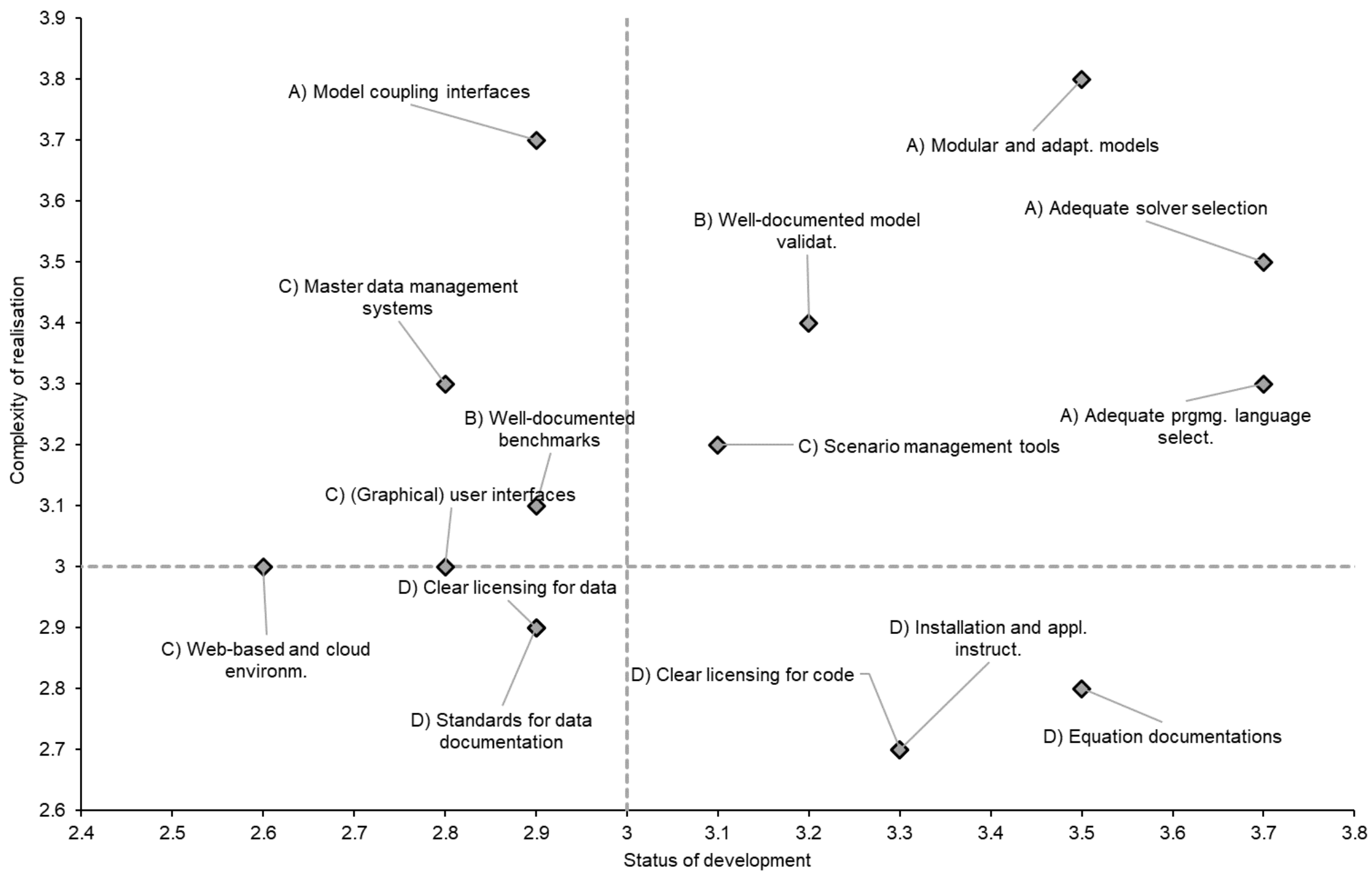

*Figure 8: Overview of the modelling strategy matrix for the average ratings of the implementation approaches regarding the Status of Development (ordinal scale very low 1- very high 5) and Complexity of Realisation (ordinal scale very low 1- very high 5) from the perspective of the whole survey sample. The categories of the methodologies are high-resolution modelling (A), programming formulations (B), model characteristics (C).*

The interrelations between the Status of Development and Complexity of Realisation (c.f. Table A3), however, also show that the realisation of especially poorly assessed items of the criterion Status of Development of category (C) might be underestimated. Experts who rate the status higher also perceive the complexity higher, as the positive and significant Spearman's rank correlation coefficients demonstrate, e.g., for the *web-based environment* (ρ=.418) or the *master data management systems* (ρ=.273). We assume again that the experts also encounter new difficulties in models with a higher degree of progress. The same applies to *well-documented benchmarks* (ρ=.311) and *scenario management tools* (ρ=.325). Moreover, statistically significant relationships between the reported model type and rated criteria are only found concerning the status item *equations documentations* (Kruskal–Wallis H test). In this context, experts from the optimisation field consider the documentation as more advanced than experts working with simulation (Mann–Whitney U test). Additionally, a pairwise comparison shows that the *modularity of the system* is assessed significantly worse from the perspective of small-scale experts than by the sample of medium- and large-scale experts.

While PhD students again rank the development status of nearly all implementation approaches lower than other respondents, the Mann–Whitney U test shows significantly lower ratings for all items of the category model usability (C) and most of the items of the category documentation standards (D). This might highlight the difficulties a PhD student has to understand and apply existing models.

### 3.5 Management items

In the final section of the survey, the experts report the lowest status ratings concerning human resources management (A). Thereby, both the existence of *continuous training* (3.1±1.3, n=54) and the possibility of *recruiting adequately trained staff* (3.2±1.1, n=51) are rated lowest on average concerning the actual development status. This is directly followed by the issues regarding compliance with *requirements for open access, open data, and open-source code* (3.2±1.3, n=55), as well as the presence of *continuous model maintenance and version control* (3.4±1.3, n=54) and *technical infrastructure* (3.5±1.1, n=56). The *availability of appropriate journals* (4.0±1.1, n=59) and the possibility of *public presentation of the project results* (3.7±1.2, n=59) are hardly seen as a current problem by all respondents. Interestingly, experts from universities rate the status of all management issues lower than experts from research institutions or companies. The highest difficulty of realisation is reported for the possibility of *recruiting adequately trained staff* (3.8±1.0, n=51) and the presence of *continuous model maintenance and version control* (3.4±1.2, n=56). In this context, respondents from universities are more confident and rated the difficulty of realisation for nearly all the issues slightly lower than institutional respondents. This is especially true regarding the difficulty of *recruiting future staff* (uni: 3.7±1.1, n=37; inst:4.6±0.5, n=9). That is surprising as one challenge of universities, in general, is to compete with industry, where good employees can typically earn much more.

## 4. Discussion

The analysed survey responses and the elaborated modelling strategy matrix demonstrate the representation states and future needs of various items in ESM. In the following, we embed the key results into the body of literature. While Section 4.1 discusses cross-cutting aspects of all survey sections, Section 4.2 and Section 4.3 discusses results related to capability and methodology aspects as well as implementation and management aspects, respectively. Finally, we present the limitations of this paper in Section 4.4.

### 4.1 Cross-cutting aspects

The field of ESA is becoming increasingly complex, which includes the models themselves as well as coupling exercises (Kotzur et al. 2020; Priesmann et al. 2019). Evidence for this is in the number of items extracted from the reviews and included in the survey, as well as the overall high complexity ratings in the results. Including more and more capabilities into the models themselves can involve trade-offs with regard to the understandability of the models. As one of the experts commented in the survey, “there is a difficult balance between simplicity and detail in energy modelling”. A solution for that problem would be, as another expert suggests, a modular approach where the user can customise just the relevant part of the model while applying pre-defined settings and data for the other sectors.

The question remains, to what extent the modeller has to and is able to fully understand the whole model. Our results show that PhD-students rate items of model usability significantly lower than more senior researchers, which might indicate that ESMs have become so complex that it is a time-consuming and demanding task to fully understand them. To improve the knowledge of the mechanisms of the model functionalities and interdependencies between input and output, simultaneous visualisation of results when changing model input was suggested by one expert. This could on the one hand help the modeller to understand the model better, but also improve the dialogue between the modeller and stakeholder/decision-maker. Indeed, Chang et al. (2021) point the latter out as one of the main current challenges of ESM.

As other research also confirms, the lack of transparency is still an issue for ESM (Junne et al. 2019; Morrison 2018) and our survey confirms that difficulties regarding compliance with requirements for open access, open data and open source are present. However, although there is a growing number of openly available models[3] (Oberle und Elsland 2019), many researchers still program and use their own ones. Regarding data, keeping consistent and high-quality data sources is seen as highly advanced but also the most complex to realise – it is still a complex and time-consuming part of the modelling work. The outlined challenges by David Stuart et al. (2018) support the results that data sharing is difficult and at the same time required data for energy models are so extensive and from so many different sources that this is actually the main part of the time-load in the modelling life. According to their survey, half of the researchers see challenges in organising data in a presentable and useful way. Furthermore, more than one-third are unsure about the copyright and licensing of the data. Additionally, one quarter also reported that there is a lack of time to share and deposit data (David Stuart et al. 2018). At the same time, there are a rising number of projects where modellers try to free time for their modelling work by sharing the data-work-load[4]. Furthermore, there has been some development regarding sharing of the tedious data work for ESM[5] and also a provision of data from official sources (e.g. ENTSO-E for electricity generation, transport and consumption data; OpenStreetMap for building and other infrastructure data), freeing time for the modellers to focus on the analyses. Thus, there are initiatives for data sharing and open-source models available, but what is lacking is – as one of the experts suggests – open source as a structured approach. This could be a data and model hub, maintained by the EU or other central administration. A good example of how this could be realised is the Danish Energy Technology Catalogue, provided by the Danish Energy Agency in collaboration with different experts[6]. A more far-reaching suggestion is a global structure (network or platform) for discussing results, stakeholder engagement, policy modelling, scenario structures and data sources for all energy system analysis/models. From a European point of view, the Energy Modelling Platform Europe[7] could be a starting point for that. Some of the current H2020-projects are contributing to that forum for exchanging research, development and practice of energy system modelling. The explicit goal of the OpenEntrance[8] project is to develop, use and disseminate an open, transparent and integrated

[3] for initial insights see https://wiki.openmod-initiative.org/wiki/Open_Models
[4] for an example see https://open-power-system-data.org/
[5] for an overview see https://wiki.openmod-initiative.org/wiki/Data; for examples for projects or platforms see https://open-power-system-data.org/; https://openenergy-platform.org/
[6] for more information see https://ens.dk/en/our-services/projections-and-models/technology-data
[7] for more information see https://www.energymodellingplatform.eu/
[8] for more information see https://openentrance.eu/

modelling platform. In the US, the National Renewable Energy Laboratory has initiated workshops on the use of open data and open software tools in the energy modelling community in North America[9]. An US-based open energy data portal is developed within OpenEI[10]. Also noteworthy are some of the open-source frameworks like PyPSA (Brown et al. 2018), oemof (Hilpert et al. 2018), Balmorel (Wiese et al. 2018a), Calliope (Pfenninger und Pickering 2018), TEMOA (DeCarolis et al. 2012) or Backbone (Helistö et al. 2019).

An obvious question arises in the context of cross-sectional items: do we need innovative new methods or rather incremental progress by combining already existing methods and ideas? In terms of novel or groundbreaking methods with the potential to revolutionise the ESA field, there was some diversity in opinion amongst the experts. Some pointed towards multi-objective and near-optimal solutions with Pareto fronts, as well as leader-follower equilibria with bi-level optimisation models. Others only referred to partial solutions with the main challenge being to integrate these in the most effective way and with an acceptable effort.

Whilst individual modelling approaches are already some of the most advanced in terms of complexity and development, there has been a strong trend towards coupling diverse models in order to exploit their respective benefits (Hansen et al. 2019; Crespo del Granado et al. 2018; Krook-Riekkola et al. 2017). Such approaches were further emphasised by the experts as continued avenues to achieve developmental advances, for example with multi-scale energy system modelling and coupling of system dynamics and optimisation models. Whilst there has been no clear answer to the above question from our survey, what definitely becomes clear is that many questions cannot be answered by models directly but are rather part of the scenario process and the way results are interpreted. Here interdisciplinarity (energy analysts, environmentalists, economists, social sciences) is essential for analyzing and questioning the framework ESM is embedded in and thus restricted to. As already pointed out decades ago, the key benefit in ESA is not the ESM itself, but the knowledge the experts gain while working with the models – the model being rather the tool to help the expert understand the system than to provide concrete answers: modelling for insights, not for numbers (Huntington et al. 1982).

## 4.2 Capability and methodology aspects

The capability results show that especially some of the wider and socioeconomic aspects of ESA can be considered *tough nuts*. This includes items like stakeholder dynamics and lifestyle aspects (with technology acceptance and adoption as a moderately developed top stars), as well as non-energy sector impacts, market design and inter-market modelling. For these topics at least, the high complexity can be understood as the main reason for a lack of development. Lower down the matrix, however, there are topics such as equity and distributional effects and material resource assessments, for which the lack of development is apparently not solely due to complexity. Instead, these research questions are related to relatively recent topics within the ESM community, which for this reason have not yet reached a high stage of development.

Socioeconomic aspects of energy systems, especially relating to behaviour, decision making and acceptance, have become more important in the ESA field recently and also feature strongly in the survey. Indeed, the originally mainly techno-economic focus of ESA in terms of energy systems and

[9] for more information see https://www.nrel.gov/analysis/open-energy-modeling-north-america-workshop.html
[10] for more information see https://openei.org/wiki/

markets continues to be extended into the social domain, especially but not only within the framework of socio-technical transitions (Li et al. 2015). In this context, equity and distributional effects have also come to the fore in recognition of the fact that energy transitions not only impact coal miners but instead have a wider and more diverse set of impacts on equally different stakeholders (Carley und Konisky 2020). But there is still a large scope for improvement in terms of the ways in which the ESA community accounts for distributional effects in its models if this is done at all. One reason for this lack of attention (at least in a European context) might be consistent metrics and datasets for energy poverty, fuel poverty and energy vulnerability, which have not been available until recent years (Thomson et al. 2017). The importance of such aspects is emphasised by one of the experts, who suggests that spatial justice could be linked to project finance and social physics techniques could be included in numerical models. Indeed, there have been some attempts to do include distributional impacts in long terms energy scenarios (Fell et al. 2020).

Several aspects of capability and methodology can be broadly interpreted in the context of widening the ESM scope or system boundary. This especially applies to material resource assessments, land-use planning patterns and non-energy sector impacts. Whilst the former two lie on the boundary between low hanging fruits and tough nuts with a complexity score of 3.0, with a score of 3.4 the latter is definitively a tough nut. Land use planning patterns are typically considered in broader ESAs such as Integrated Assessment Models (IAMs) and are particularly relevant where questions relating to agricultural land for food, chemicals and/or energy are posed. Especially where net-zero scenarios are being explored with bioenergy with carbon capture and storage, Gambhir et al. (2019) conclude that IAMs benefit most from couplings with other models and approaches. With the increase in modern bioenergy exploitation in recent decades, the relevance of land-use competition issues for these applications has also come to the fore. In addition, there is a trend in the ESA community towards combining ESMs with LCA methods, in order to account for the impact of energy technologies beyond their operational phase (Thomson et al. 2017). But such endeavours present several challenges, including different temporal horizons or system boundaries, data quality and availability, and the underrepresentation of industrial processes (Kullmann et al. 2021). In terms of non-energy sector impacts, one of the experts suggested studying the investments (or opportunities) outside the energy sector, which might explain some lack of investment in energy-related infrastructure for energy efficiency.

Related to the above-mentioned challenges of stakeholder dynamics and lifestyle aspects, simulation models are viewed as significantly more advanced than optimisation models to represent technology acceptance and adoption by sub-groups who work with optimisation or simulation models. For example, two separate experts highlighted the combination of agent-based methodologies (ABMs) to study consumer preferences and technology diffusion with optimisation modelling to analyse optimal technology paths for future energy systems. In contrast, optimisation models are significantly more developed concerning ramping capabilities, detailed technology process models, and supply-side flexibility options according to the modellers using them. Furthermore, a higher average status rating is reported for smaller than larger-scale modelling experts for the categories social aspects and human behaviour modelling and transmission and distribution system modelling. In contrast, flexibility, sector coupling and energy system integration modelling is rated on average slightly higher by large scale modelling experts. These findings are in line with the features of simulation and optimisation models as such. Optimisation models are good when relationships can be described in simple, often linear terms, thus are well suited especially for the supply side modelling and supply-side flexibility (Lund et al. 2017;

Epelle und Gerogiorgis 2020). Since simulation model have fixed assumed capacities and can be though as if-then decisions in ESM, they are capable of taking into account more complex, often non-linear relationships, which makes them a good choice for demand-side modelling, including for demand-side flexibility. This also includes modelling acceptance and adoption, as well as end-user behaviour (Lund et al. 2017).

Furthermore, simulation models are well suited for modelling different approaches, which then ask for more active involvement of stakeholders in the decision-making process (Bruckner et al. 2005). However, detailed representations of demand-side modelling with complex non-linear relationships, as well as the vast amount of input data needed, makes it burdensome to apply the same methodologies to the large-scale models. As pointed out by one of the respondents, a combined analysis of sociological and technological dynamics might be helpful to assess transformation pathways more realistically by providing insights into the interactions between the decision processes of market actors and the performance of the supply system. Such approaches have been followed, e.g. with empirically grounded agent-based modelling and optimisation models by Wittmann (2008), Chappin und Dijkema (2010) or Scheller et al. (2018).

Many experts also mentioned combinatorial optimisation approaches (e.g. graph theory) or machine learning (ML) as important future research methods in the field of ESA. The energy research field is indeed one of the most important areas for which combinatorial optimisation methods are applied and developed today (Weinand et al. 2021), e.g., for optimal power flow planning (Abido 2002) or designing of district heating networks (Weinand et al. 2019). However, the underlying combinatorial problems are often NP-hard, i.e. very difficult to solve exactly (Goderbauer et al. 2019). ML-based approaches which show promising results in different applications by making decisions that were otherwise made by handcrafted expert knowledge-based heuristics in a more principled and optimised way could help to solve these problems (Bengio et al. 2021). In a recent collaborative study of some of the most important experts of the ML community, ML methods to tackle climate change have been proposed (Rolnick et al. 2019). The article contains a compilation of ML methods, which could be used for various problems like "optimising buildings", "urban planning" or "modelling social interactions". Specific examples, include designing energy systems (Perera et al. 2019), determining long-term dependencies in occupant behaviour (Kleinebrahm et al. 2021) or price forecasting in electricity market simulations (Fraunholz et al. 2021).

## 4.3 Implementation and management aspects

The most important implementation items are related to the usability and documentation of models as well as model modularity, whereas the management items are focused on requirements for open access, open data, and open-source code and the recruitment of adequate staff. While various items regarding model usability and documentation standards have scope for improvement, experts from the optimisation field considered the equation documentations significantly more advanced than experts working with simulation models. This is probably due to the constraints imposed by employing an optimisation model, whereby the model should have a pre-defined structure and concerns about solvability and run times may lead to more rigorous documentation. Simulation-based approaches, on the other hand, arguably offer more freedom for experimentation, with less of a clearly defined structure and objective (Möst et al. 2009). Higher usability might be achieved by adding data management systems and graphical user interfaces to the existing models.

The high-status ratings but also the recognition of the complexity of items such as the modularity and adaptability of models as well the adequate solver selection suggest strong activity in the field over the past decade(s) (Pfenninger et al. 2014). Thereby, the actual modularity of the system is rated significantly lower from the perspective of small-scale experts and higher from long-term experts. This may show that the diversity in research questions to be answered has increased, whereby large ESMs are required to be modular to remain feasible. This is also related to adaptability, whereby a model should be easily tailored (and tailorable) to a specific research question or application. When it comes to documentation of models and data, it seems that the former is considered more advanced than the latter. In other words, aspects such as master data management systems, well-documented benchmarks, graphical user interfaces, clear licensing for data and standards for data documentation all exhibit lower than average levels of development. This goes in line with the suggestion of Keirstead et al. (2012). On the other hand, the model- and code-related aspects appear on the right-hand side of Figure 8. This may reflect advances in making models transferable, open-source and/or validated, all with good supporting documentation, but which is lacking for their data framework (Mendes et al. 2011).

In terms of the management items, PhD students reported lower ratings for model usability and documentation standards than more senior researchers. This may highlight the difficulties of young researchers applying models. In addition, since professors even indicate a significantly higher status in data documentation standards, this may indicate that particular tasks might be underestimated in terms of their complexity. Indeed, PhD projects in the ESA field often invest large amounts of effort to get a full picture of the model. Probably this also relates to stepwise model developments over longer periods of time, each adding additional layers of complexity, whereby the senior scientist(s) and/or group leaders are the only ones who still have (or are still able to keep) an overview.

## 4.6 Study limitations

An inherent limitation of survey-based research is that respondents may assess their own perceptions, differently in different contexts . For instance, a tendency exists to assess one's own research field more positively or more complex in such a public setting (Graeff 2005; Bogner und Landrock 2016). This also might contradict our assumption in terms of the correlations between the two evaluation criteria that experts who directly work with these approaches experienced a higher problem complexity with each advancement. Although the compilation and classification of the queried items in terms of the different survey sections are derived from a comprehensive review process, we might have missed relevant items. Furthermore, our terms for the items might be understood in different ways by the different experts. Since it is challenging to agree on a specific vocabulary for all respondents, we expressed our understanding of each item with an additional definition included in the survey.

Related to the empirical design, we prioritised the expert knowledge of respondents over the number of respondents. The sample size and bias is quite small. Nevertheless, the respondents of the sample cover various countries of 20 most productive countries in the field of ESA (Dominković et al. 2021). Furthermore, our sample size is comparable to similar studies (Chang et al. 2021; Connolly et al. 2010). While there are even more answers available for optimisation models, the small size is particularly visible for all other models. In this regard, the determination of statistically significant relationships between the reported model type and rated criteria should be taken with caution. The same applies to other modelling sub-groups such as temporal and spatial scales. Additionally, the survey does not sufficiently allow for interdisciplinary studies since each respondent needed to choose a single model type to rate the items, as one respondent correctly remarked. For instance, energy analysts working alongside environmentalists, economists and social scientists using soft-linked models can hardly assess the questionnaire from their comprehensive view. Nevertheless, we could hardly find any significant differences in the context of the average ratings between experts of different modelling types. While this made it even more difficult to clearly identify future modelling challenges and opportunities, our specially-created matrix revealed valuable insights by comparing the items regarding the two criteria and classifying them in the different quadrants.

# 5. Summary and conclusion

This paper seeks to contribute to the literature on future research opportunities and challenges for ESA. For this, we conducted a quantitative expert survey with a sample size of N=61 to provide insights regarding the criteria *Status of Development* and the *Complexity of Realisation* of 96 identified and classified question or rather modelling items from various reviews in the ESA and ESM field. With the two criteria in mind, a specially defined 2x2 modelling strategy matrix is applied to determine modelling items that are poorly developed and easy to implement ("*low hanging fruits*"), poorly developed and complex to implement ("*tough nuts*"), highly developed and easy to implement ("*long runners*"), and highly developed and complex to implement ("*top stars*"). The expert survey does not show precise results regarding the main challenges for ESM. Although there are tendencies for *low hanging fruits* and *tough nuts*, there are hardly any outliers. In more detail, we identify capabilities like land-use planning patterns, equity and distributional effects and endogenous technological learning as *low hanging fruits* for enhancement and a large number of complex topics that are already well implemented. The remaining *tough nuts* regarding modelling capabilities include non-energy sector and social behaviour interaction effects.

The general level of complexity in the field of energy system modelling is rather high as well as the diversity of modellers, model types and applications. Instead of converging model types, the combination of advantages of model techniques by model coupling is high on the agenda. However, this further increases the complexity of result interpretation and the respective result communication, which has the potential to become a research field on its own. Considering the already high-level complexity of many models, the true art seems to be choosing a manageable level of complexity instead of adding up to the already existing complexity. In general, openness is a way forward and on top of transparency, accessibility of models and collaboration could open the way for more interdisciplinary ESA, which can combine the specialities of the different modelling techniques and types.

# Acknowledgements

Fabian Scheller kindly acknowledges the financial support of the European Union's Horizon 2020 research and innovation programme under the Marie Sklodowska-Curie grant agreement no. 713683 (COFUNDfellowsDTU). The authors would also like to thank all of the survey participants for their time and expertise in completing the survey.

# Appendix A: Overview of survey response

*Table A1: Average rating of the modelling capabilities regarding the Status of Development (ordinal scale very low 1- very high 5) and Complexity of Realisation (ordinal scale very low 1- very high 5; based on the scale a colour transition from red over white to blue or rather lower ratings over medium ratings to higher ratings is applied in this table) from the perspective of the whole survey sample (N) as well of the sub-sample of optimisation model users (Opt), simulation model users (Sim), and other model users (Ors). The modelling capabilities are sorted in ascending order from the perspective of the whole survey sample (capabilities with the lowest Status of Development are at the top). The pairwise Spearman coefficient (ρ) between the rating of the Status of Development and the Complexity of Realisation is also presented for each of the groups (* correlation is significant at the 10% level). The categories of the capabilities are social aspects and human behaviour modelling (A), demand-side modelling (B), transmission and distribution system modelling (C), supply generation modelling (D), flexibility, sector coupling and energy system integration modelling (E), markets and regulations framework modelling (F), environmental and resources modelling (G), as well as feedback and interaction effects (H). The numbers (1-32) are related to the question items of the modelling capabilities (c.f. Table 1).*

| Capability items | | | Status of development | | | | Complexity of realisation | | | | rho (Status, Complexity) | | | |
|---|---|---|---|---|---|---|---|---|---|---|---|---|---|---|
| | | | N | Opt | Sim | Ors | N | Opt | Sim | Ors | N | Opt | Sim | Ors |
| A | 2 | Lifestyle aspects | 2.3 | 2.1 | 2.2 | 2.6 | 3.2 | 3.0 | 3.6 | 3.3 | -.025 | -.006 | -.494 | .342 |
| A | 5 | Equity & distributional effects | 2.3 | 2.2 | 2.0 | 2.6 | 3.1 | 2.8 | 4.0 | 2.9 | -.005 | -.165 | -.197 | .930* |
| F | 25 | Market design | 2.3 | 2.3 | 2.4 | 2.2 | 3.3 | 2.9 | 4.1 | 3.4 | .135 | .047 | .447 | .046 |
| H | 32 | Non-energy sector impacts | 2.3 | 2.2 | 2.2 | 2.8 | 3.4 | 3.2 | 4.0 | 3.0 | -.058 | -.018 | -.458 | .379 |
| G | 27 | Land-use planning patterns | 2.4 | 2.5 | 1.8 | 2.8 | 3.0 | 2.8 | 3.7 | 3.0 | .175 | .148 | .082 | .687 |
| A | 3 | Stakeholder dynamics | 2.4 | 2.3 | 2.7 | 2.8 | 3.4 | 2.8 | 4.0 | 3.7 | -.011 | -.119 | .013 | .151 |
| C | 13 | Virtual power plants | 2.6 | 2.7 | 2.1 | 2.6 | 3.0 | 2.8 | 3.6 | 3.0 | .327* | .445* | .690* | .000 |
| H | 30 | Endogenous techn. learning | 2.6 | 2.4 | 2.7 | 3.4 | 3.1 | 2.8 | 4.1 | 3.1 | -.023 | .006 | -.041 | -.163 |
| G | 29 | Nexus issues | 2.6 | 2.5 | 2.5 | 3.3 | 3.3 | 3.3 | 4.0 | 3.1 | -.123 | -.209 | -.099 | .194 |
| C | 11 | Gas network characteristics | 2.6 | 2.5 | 2.5 | 3.1 | 3.1 | 2.9 | 3.6 | 3.2 | .189 | .117 | .000 | .658* |
| F | 24 | Inter-market modelling | 2.7 | 2.7 | 2.3 | 2.9 | 3.3 | 3.1 | 4.0 | 3.4 | -.028 | -.043 | .689 | -.595* |
| G | 28 | Material resource assessments | 2.7 | 2.8 | 2.3 | 2.7 | 3.0 | 3.0 | 3.4 | 2.9 | -.063 | -.032 | -.436 | .422 |
| C | 9 | Microgrid & autonomy aspects | 2.7 | 2.8 | 2.6 | 2.7 | 3.0 | 2.8 | 3.4 | 3.3 | .100 | .109 | -.236 | .460 |
| F | 26 | Regulatory & policy frameworks | 2.8 | 2.7 | 2.2 | 3.8 | 3.2 | 3.0 | 4.0 | 3.3 | .081 | .148 | -.107 | .000 |
| H | 31 | Elastic demands | 2.8 | 2.6 | 2.3 | 3.8 | 3.2 | 2.9 | 4.1 | 3.3 | .092 | .218 | -.148 | -.150 |
| C | 14 | Ancillary services | 2.9 | 3.1 | 2.5 | 2.6 | 3.0 | 3.1 | 3.7 | 2.8 | .398* | .438* | .203 | .660* |
| E | 21 | Innovative storage modelling | 3.0 | 3.2 | 2.5 | 2.9 | 3.4 | 3.2 | 4.1 | 3.5 | .100 | .150 | -.318 | .484 |
| B | 8 | Consumption process models | 3.0 | 3.0 | 2.9 | 3.3 | 3.3 | 3.1 | 3.7 | 3.6 | .007 | .338* | -.707* | -.396 |
| A | 1 | Technol. acceptance & adoption | 3.1 | 2.8 | 3.7 | 3.4 | 3.3 | 3.2 | 3.8 | 3.3 | .403* | .270 | -.328 | .847* |
| E | 23 | Demand-side flexibility options | 3.1 | 3.2 | 2.6 | 3.1 | 3.6 | 3.4 | 4.1 | 3.5 | .046 | .112 | -.316 | .414 |
| C | 12 | Heat network characteristics | 3.2 | 3.2 | 3.1 | 3.3 | 3.2 | 3.2 | 3.3 | 2.9 | .282* | .170 | .000 | .771* |
| C | 10 | Power network characteristics | 3.2 | 3.4 | 2.8 | 3.1 | 3.3 | 3.2 | 3.7 | 3.6 | .236* | .388* | -.096 | .308 |
| D | 15 | Ramping capabilities | 3.3 | 3.6 | 2.5 | 3.2 | 3.0 | 3.0 | 3.6 | 2.9 | .007 | .109 | -.041 | .133 |
| E | 22 | Supply-side flexibility options | 3.4 | 3.6 | 2.8 | 3.2 | 3.4 | 3.3 | 4.1 | 3.4 | .038 | .226 | -.349 | -.203 |
| D | 16 | Detailed techn. process models | 3.4 | 3.7 | 2.9 | 2.7 | 3.1 | 3.0 | 3.6 | 3.1 | -.025 | -.041 | .334 | .140 |
| E | 20 | Multi-energy services & carriers | 3.5 | 3.7 | 2.9 | 3.4 | 3.4 | 3.3 | 4.1 | 3.5 | .055 | .087 | .010 | .179 |
| D | 18 | Non-conv. energy supply sources | 3.5 | 3.7 | 3.5 | 3.0 | 3.1 | 3.1 | 3.5 | 3.2 | .135 | .261 | -.078 | .149 |
| A | 4 | Techn. diffusion | 3.5 | 3.5 | 3.3 | 3.8 | 2.9 | 2.9 | 3.6 | 3.5 | .205 | .247 | -.010 | .351 |
| B | 7 | Demand-side tech. heterogeneity | 3.5 | 3.7 | 3.1 | 3.4 | 3.3 | 3.3 | 3.8 | 3.3 | .113 | .348* | -.066 | -.121 |
| E | 19 | Cross-sectoral approaches | 3.6 | 3.8 | 3.0 | 3.4 | 3.5 | 3.3 | 4.2 | 3.6 | -.035 | .006 | -.199 | .109 |
| B | 6 | Energy service demands | 3.7 | 3.6 | 4.0 | 3.5 | 3.2 | 3.1 | 3.4 | 3.4 | -.087 | -.072 | -.139 | .113 |
| D | 17 | Supply-side techn. heterogeneity | 3.7 | 3.9 | 3.1 | 3.6 | 3.0 | 3.1 | 3.2 | 3.2 | .052 | .145 | -.054 | -.059 |

*Table A2: Average rating of the methodological approaches regarding the Status of Development (ordinal scale very low 1- very high 5) and Complexity of Realisation (ordinal scale very low 1- very high 5; based on the scale a colour transition from red over white to blue or rather lower ratings over medium ratings to higher ratings is applied in this table) from the perspective of the whole survey sample (N) as well of the sub-sample of optimisation model users (Opt), simulation model users (Sim), and other model users (Ors). The modelling methodologies are sorted in ascending order from the perspective of the whole survey sample (methodologies with low Status of Development are at the top). The pairwise Spearman coefficient (ρ) between the rating of the Status of Development and the Complexity of Realisation is also presented for each of the groups (* correlation is significant at the 0.1 level). The categories of the methodologies are high-resolution modelling (A), programming formulations (B), model characteristics (C). The numbers (1-15) are related to the question items of the methodological approaches (c.f. Table 1).*

| Methodology items | | | Status of development | | | | Complexity of realisation | | | | rho (Status, Complexity) | | | |
|---|---|---|---|---|---|---|---|---|---|---|---|---|---|---|
| | | | N | Opt | Sim | Ors | N | Opt | Sim | Ors | N | Opt | Sim | Ors |
| B | 6 | New general mathem. frameworks | 2.8 | 2.8 | 2.8 | 2.8 | 3.4 | 3.1 | 4.1 | 3.3 | .289* | .407* | .197 | .118 |
| B | 7 | NLP formulations | 2.9 | 2.7 | 3.3 | 3.1 | 3.8 | 3.9 | 3.9 | 3.6 | .189 | .152 | .159 | .627 |
| A | 4 | Decomposition methods | 2.9 | 3.0 | 2.7 | 2.9 | 3.6 | 3.4 | 4.1 | 3.8 | .167 | .241 | .049 | -.015 |
| C | 12 | Focus on uncertainty analysis | 2.9 | 2.9 | 2.5 | 3.2 | 3.4 | 3.2 | 4.1 | 3.4 | .280* | .312* | .005 | .466 |
| B | 10 | SP formulations | 2.9 | 2.9 | 2.5 | 3.6 | 3.3 | 3.3 | 3.6 | 3.4 | .328* | .481* | -.310 | .425 |
| A | 3 | Foresight approaches | 3.0 | 3.2 | 2.6 | 3.0 | 3.4 | 3.3 | 4.1 | 3.3 | .043 | .038 | -.093 | .338 |
| A | 5 | Soft- or hard-coupling of models | 3.1 | 3.2 | 2.9 | 3.1 | 3.7 | 3.6 | 4.3 | 3.8 | .174 | .141 | .453 | .028 |
| C | 13 | Sustainability indicator assmnt. | 3.1 | 3.1 | 3.0 | 3.3 | 3.1 | 2.9 | 3.8 | 3.1 | .031 | .033 | .021 | -.072 |
| C | 15 | Integ. assmnt. multi capabilities | 3.2 | 3.4 | 2.8 | 2.9 | 3.5 | 3.5 | 4.0 | 3.4 | .070 | .075 | .098 | -.099 |
| A | 1 | High(er) level of spatial disag. | 3.3 | 3.2 | 3.1 | 3.5 | 3.3 | 3.2 | 3.9 | 3.2 | .077 | .185 | -.143 | -.140 |
| A | 2 | High(er) level of temporal disag. | 3.5 | 3.6 | 3.2 | 3.5 | 3.3 | 3.1 | 4.0 | 3.3 | .132 | .223 | .577 | -.235 |
| C | 14 | Technology neutrality | 3.6 | 3.7 | 3.2 | 3.6 | 3.2 | 3.1 | 3.4 | 3.1 | .088 | .094 | .085 | .190 |
| C | 11 | Consistent quality data sources | 3.7 | 3.7 | 3.7 | 3.5 | 3.8 | 3.9 | 4.1 | 3.4 | .099 | .074 | -.185 | .470 |
| B | 8 | MIP formulations | 3.7 | 3.9 | 3.1 | 3.6 | 3.5 | 3.5 | 3.4 | 3.3 | .384* | .318* | .559 | .372 |
| B | 9 | LP formulations | 4.0 | 4.3 | 3.1 | 3.8 | 3.1 | 2.9 | 3.5 | 3.2 | -.026 | .020 | .029 | .014 |

*Table A3: Average rating of the implementation approaches regarding the Status of Development (ordinal scale very low 1- very high 5) and Complexity of Realisation (ordinal scale very low 1- very high 5; based on the scale a colour transition from red over white to blue or rather lower ratings over medium ratings to higher ratings is applied in this table) from the perspective of the whole survey sample (N) as well of the sub-sample of optimisation model users (Opt), simulation model users (Sim), and other model users (Ors). The modelling methodologies are sorted in ascending order from the perspective of the whole survey sample (methodologies with the lowest Status of Development are at the top). The pairwise Spearman coefficient (ρ) between the rating of the Status of Development and the Complexity of Realisation is also presented for each of the groups (* correlation is significant at the 0.1 level). The categories are related to development activities (A), model validation and benchmarking (B), model usability (C), and documentation standards (D). The numbers (1-15) are related to the question items of the implementation approaches (c.f. Table 1)*

| Implementation items | | | Status of development | | | | Complexity of realisation | | | | rho (Status, Complexity) | | | |
|---|---|---|---|---|---|---|---|---|---|---|---|---|---|---|
| | | | N | Opt | Sim | Ors | N | Opt | Sim | Ors | N | Opt | Sim | Ors |
| C | 9 | Web-based & cloud environm. | 2.6 | 2.5 | 2.6 | 2.7 | 3.0 | 2.8 | 3.7 | 3.1 | .418* | .498* | .424 | .407 |
| C | 10 | Master data mgnmt. systems | 2.8 | 2.8 | 2.8 | 2.7 | 3.3 | 3.2 | 3.9 | 3.0 | .273* | .520* | -.121 | .000 |
| C | 7 | (Graphical) user interfaces | 2.8 | 2.8 | 2.8 | 3.0 | 3.0 | 2.9 | 3.2 | 3.1 | .165 | .278 | -.224 | .289 |
| D | 15 | Clear licensing for data | 2.9 | 3.0 | 2.8 | 2.6 | 2.9 | 2.8 | 2.8 | 3.6 | .074 | .101 | .124 | -.210 |
| D | 13 | Standards for data document. | 2.9 | 2.9 | 2.9 | 2.8 | 2.9 | 2.8 | 2.7 | 3.2 | .001 | -.206 | .418 | .270 |
| A | 4 | Model coupling interfaces | 2.9 | 2.9 | 2.8 | 3.0 | 3.7 | 3.6 | 3.8 | 4.1 | .045 | .134 | -.246 | .103 |
| B | 6 | Well-documented benchmarks | 2.9 | 3.0 | 3.0 | 2.6 | 3.1 | 3.0 | 3.4 | 3.1 | .311* | .254 | .480 | .448 |
| C | 8 | Scenario management tools | 3.1 | 3.2 | 3.1 | 2.6 | 3.2 | 3.1 | 3.6 | 3.1 | .325* | .457* | .215 | -.308 |
| B | 5 | Well-documented model validat. | 3.2 | 3.2 | 3.0 | 3.3 | 3.4 | 3.5 | 3.7 | 3.1 | .150 | .033 | .521* | .187 |
| D | 14 | Clear licensing for code | 3.3 | 3.4 | 2.8 | 3.5 | 2.7 | 2.5 | 2.8 | 3.1 | .080 | .115 | .124 | .063 |
| D | 11 | Installation & applic. Instruction | 3.3 | 3.5 | 2.8 | 3.3 | 2.7 | 2.5 | 2.6 | 3.4 | .139 | .210 | .435 | -.538 |
| A | 3 | Modular & adaptable models | 3.5 | 3.6 | 3.3 | 3.6 | 3.8 | 3.7 | 3.7 | 4.1 | .191 | .138 | .160 | .493 |
| D | 12 | Equation documentations | 3.5 | 3.8 | 3.0 | 3.4 | 2.8 | 2.7 | 2.5 | 3.4 | .121 | .067 | .185 | .455 |
| A | 2 | Adequate solver selection | 3.7 | 3.8 | 3.4 | 3.6 | 3.5 | 3.4 | 3.9 | 3.6 | .115 | .282 | -.543 | .000 |
| A | 1 | Programming. Language select. | 3.7 | 3.7 | 4.0 | 3.4 | 3.3 | 3.4 | 2.9 | 3.7 | -.053 | .064 | -.642* | .674* |

# Appendix B: Supplementary material

The Supplementary Material (SM) consists of the results of the literature analysis concerning the research scope and future needs (SM A), the complete overview of the survey questions (SM B), and the cover letter of the invitation mail explaining the intention and background of the study (SM C).